%% file: paper.tex
\documentclass[sigconf]{acmart}

\usepackage{amsmath,amsfonts}
\usepackage{algorithmic}
\usepackage{graphicx}
\usepackage{textcomp}
\usepackage{xcolor}
\usepackage{subfiles}
\usepackage{tabularx}
\usepackage{booktabs}
\usepackage[capitalise]{cleveref}
\usepackage{xspace}
\usepackage{url}
\usepackage{caption}
\usepackage{subcaption}
\usepackage{balance}
\usepackage{tabularx}
\usepackage{bm}
\usepackage{paralist}

\copyrightyear{2024}
\acmYear{2024}
\setcopyright{acmlicensed}\acmConference[ICSE '24]{2024 IEEE/ACM 46th International Conference on Software Engineering}{April 14--20, 2024}{Lisbon, Portugal}
\acmBooktitle{2024 IEEE/ACM 46th International Conference on Software Engineering (ICSE '24), April 14--20, 2024, Lisbon, Portugal}
\acmPrice{15.00}
\acmDOI{10.1145/3597503.3623339}
\acmISBN{979-8-4007-0217-4/24/04}

\graphicspath{{./}{./img/}}

\newcommand{\summary}[2]{%
	\vspace{-0.2cm}%
	\begin{center}%
		\colorbox{gray!20}{%
			\parbox{\linewidth}{%
				\textbf{\textsf{Summary (\textit{#1})}:}~%
				#2%
			}%
		}%
	\end{center}%
}

\newcommand{\toolname}{\emph{\mbox{IntelliGame}}\xspace}
\newcommand{\passau}{University of Passau\xspace}
\newcommand{\control}{\emph{control}\xspace}
\newcommand{\notifications}{\emph{notifications}\xspace}
\newcommand{\treatment}{\emph{treatment}\xspace}
\newcommand{\maximizing}{\emph{maximizing}\xspace}

\makeatletter
\newcommand\footnoteref[1]{\protected@xdef\@thefnmark{\ref{#1}}\@footnotemark}
\makeatother

\acmConference[ICSE 2024]{46th International Conference on Software Engineering}{April 2024}{Lisbon, Portugal}


\begin{document}

\title{Influencing Testing Behavior Using Gamification}
\title{Gamifying Testing in the Developers' IDE}
\title{Incentivizing Developers to Test within their IDE}
\title{Encouraging Developers to Test by Gamifying their IDE}
\title{Encouraging Developers to Test using Gamification}
\title{Encouraging Developers to Test using IDE Gamification}
\title{Gamifying Testing in the IDE to Encourage Developers}
\title{Gamifying Testing within Developers' IDEs}
\title{Gamifying Testing within Developers' IDEs}
\title{AchiJ: Gamifying Software Testing in IntelliJ}
\title{Gamifying Software Testing in IntelliJ}
\title{Encouraging Developers to Test by Gamifying their IDE}
\title{Encouraging Developers to Test by Gamifying IntelliJ}
\title{Encouraging Developers to Test using IDE Gamification}
\title{Improving Testing Behavior in IntelliJ using Gamification}
\title{Improving Testing Behavior by Gamifying IntelliJ}

\author{Philipp Straubinger}
\affiliation{%
	\institution{University of Passau}
	\city{Passau}
	\country{Germany}
}

\author{Gordon Fraser}
\affiliation{%
	\institution{University of Passau}
	\city{Passau}
	\country{Germany}
}


\begin{abstract}
  Testing is an important aspect of software development, but
  unfortunately, it is often neglected. While test quality analyses
  such as code coverage or mutation analysis inform developers about
  the quality of their tests, such reports are viewed only
  sporadically during continuous integration or code review, if they
  are considered at all, and their impact on the developers' testing
  behavior therefore tends to be negligible. To actually influence
  developer behavior, it may rather be necessary to motivate
  developers directly within their programming environment, while they
  are coding.
  We introduce \toolname, a gamified plugin for the popular IntelliJ
  Java Integrated Development Environment, which rewards developers
  for positive testing behavior using a multi-level achievement
  system: A total of 27 different achievements, each with incremental
  levels, provide affirming feedback when developers exhibit commendable
  testing behavior, and provide an incentive to further continue and
  improve this behavior.
  A controlled experiment with 49 participants given a Java
  programming task reveals substantial differences in the testing
  behavior triggered by \toolname: Incentivized developers write more tests, achieve higher coverage and mutation
  scores, run their tests more often, and achieve functionality earlier.
\end{abstract}

\maketitle

\section{Introduction}

\subfile{sections/introduction}

\section{Background}

\subfile{sections/background}

\section{Gamification Plugin for IntelliJ}

\subfile{sections/plugin}

\section{Evaluation}

\subfile{sections/evaluation}

%\section{Results}

\subfile{sections/results}

\section{Conclusions}

\subfile{sections/conclusion}

\section{Acknowledgments}
We would like to thank Jonas Lerchenberger for the initial implementation of \toolname and his help during the experiments. This work is supported by the DFG under grant \mbox{FR 2955/2-1}, ``QuestWare: Gamifying the Quest for Software Tests''.

\balance
\bibliographystyle{ACM-Reference-Format}
\bibliography{bib}

\end{document}

%% file: sections/introduction.tex
	Although the importance of software testing is well established,
	unfortunately so is the fact that software is nevertheless generally
	inadequately tested~\cite{seth2014organizational,pooreport}.  A common approach to ensure and
	improve test quality is to measure and visualize established metrics
	such as code coverage~\cite{DBLP:journals/csur/ZhuHM97} or mutation
	analysis~\cite{5487526}. These analyses provide insights into how
	thoroughly a program is tested, and where there are holes in the test suite
	that could be improved with further tests. A common way these analyses
	are integrated into the developer workflow is by producing detailed
	reports during continuous integration or code
	review~\cite{DBLP:conf/sac/AhmadLS21,ivankovic2019code}. While such reports are popular, there is little
	evidence \cite{HERMSEN201661,Alebeisat18,MISHRA2020100308,fagerholm2015software} that showing this information has any effects on test quality
	or developer behavior.

	While it is safe to assume that software developers are nowadays well
	educated on the importance of testing and appropriate techniques to do
	so~\cite{DBLP:journals/jss/GarousiRLA20}, it appears the problem is rather one of motivation:
	Despite the availability of coverage and mutation reports, testing
	is often not tightly integrated into the developer
	workflow~\cite{DBLP:journals/tse/BellerGPPAZ19} or simply not used. In order to improve
	test quality, it may therefore be necessary to fundamentally
	influence the testing behavior of developers directly while they are
	writing code. An approach that is commonly used in many domains to
	influence the behavior of humans is gamification, i.e., the
	integration of game elements into non-game contexts to
	incentivize humans to do things they are aware of but not sufficiently
	motivated to do. While the gamification of software testing has been
	investigated previously~\cite{DBLP:conf/icse/BarretoF21, DBLP:conf/sast/JesusFPF18, robson2015all, DBLP:conf/icse/StraubingerF22, DBLP:conf/sera/Parizi16, DBLP:conf/icse/ScherrEH18}, this has not been done directly
	within the development environment to the best of our knowledge.

	In this paper, we therefore introduce an approach to
	incentivize developers to test more and better, by gamifying
	aspects of testing directly within their development
	environment, while coding. In particular, we create a catalog
	of 27 individual motivational aspects related to test quality,
	the use of test automation infrastructure, and the way testing
	activities are integrated into the workflow. Positive behavior
	according to these aspects, like measuring coverage when running tests, is rewarded by awarding developers
	multi-level \emph{achievements}, which are designed
		incrementally such that initial positive reinforcement is
		achieved quickly, while there remains an incentive to further
		improve testing behavior to reach higher achievement levels. Achievements are therefore suited for supporting onboarding activities as well as for influencing the routines of more seasoned developers. We
	implemented this approach and the 27 achievements as a plugin
	for the popular IntelliJ Java Integrated Development
	Environment (IDE), and empirically study how it influences the
	workflow of software developers.
	
	In detail, the contributions of this paper are as follows:
	\begin{compactitem}
		\item We propose a gamification approach to incentivize and reward
		testing-related behavior while coding.
		\item We introduce the \toolname tool that implements this approach with 27
		multi-level achievements integrated directly into the popular
		IntelliJ development environment.
		\item We empirically study the effects of \toolname using a
		controlled experiment with four different configurations and 49 participants tasked to implement a
		Java class. %The configurations aim to separate the effects of different parts of \toolname from each other.
	\end{compactitem}
	We observe substantial improvements:
		%differences %in the testing
		% behavior
		% triggered by \toolname:
		Incentivized developers write more tests, achieve higher coverage and mutation scores, run their tests more often, and achieve functionality earlier.
	%Notifications as stand-alone feature show increased testing behavior as well while aiming to solve as many achievements as possible leads to exploitation of certain achievements.

%% file: sections/background.tex
	\subsection{Developer Testing}
	
	Testing is an important aspect of software development, %Among many
	%different strategies and types of testing,
	and developers are
	typically expected to write unit tests for the code they
	produce~\cite{spillner2019basiswissen}. Although this is an accepted
	practice and developers often claim to spend substantial amounts of
	time on writing tests~\cite{DBLP:journals/tse/BellerGPPAZ19}, it has been shown that they tend to
	overestimate how much testing they do~\cite{DBLP:journals/tse/BellerGPPAZ19}; indeed, there are many
	developers who do not test at all, and software projects without
	automated tests~\cite{DBLP:conf/sigsoft/BellerGPZ15,DBLP:journals/tse/BellerGPPAZ19}.

	In order to help testers improve their testing, a common
		approach lies in assessing test quality using metrics like
		code coverage or mutations analysis. Code coverage can be
		easily measured by executing a test suite and checking which
		parts of the code have been reached by the tests; there are
		various definitions of what are relevant `parts', such as
		statements, branches, or execution paths. Developers receive
		feedback in terms of an overall percentage of how much of the
		program has been tested, or visualizations
		thereof~\cite{DBLP:journals/csur/ZhuHM97}.  Mutation analysis
		offers a stronger criterion by checking how many artificial
		faults can be detected by a test
		suite~\cite{5487526}. However, despite the existence of such
		means of support, software quality tends to be inadequate,
		leading to substantial costs during development and after
		releasing the software~\cite{pooreport}, suggesting that this
		feedback on test quality alone is not sufficient to motivate
		developers to test.

	\subsection{Developer Motivation}
	
	 The suspected reasons for inadequate testing are
		manifold and include ubiquitous problems such as costs and
		time, but interestingly also developer
		motivation~\cite{kapur2017release}, which is generally seen
		as an essential factor for developer
		productivity~\cite{DBLP:conf/esem/FrancaSS14,
			francca2014theory, 8370133}.  The term \emph{motivation}
		refers to either the inner willingness to participate in an
		activity for personal satisfaction (\emph{intrinsic
			motivation}), or an external outcome that comes with or
		after completing a task, such as
		recognition~\cite{ryan2000self} (\emph{extrinsic
			motivation}). Although the terms \emph{motivation} and
		\emph{satisfaction} are often used interchangeably, they are
		different~\cite{francca2014theory,
			DBLP:conf/esem/FrancaSS14}: Motivation needs to be present
		before the work, while satisfaction is a result of the
		work. However, they are connected because being satisfied
		with a task can motivate for the next one. When applied for
		a longer time, motivation can lead to \emph{engagement},
		which refers to commitment, hard work, and interest in one's
		current work and involves putting in more effort than
		required~\cite{8370133, suff2008going, devi2009employee}.

		Prior
		work~\cite{DBLP:conf/esem/SantosMCSCS17,DBLP:journals/infsof/DeakSS16}
		has investigated reasons why motivation is often missing when
		it comes to software testing. It has been found that software
		testing is often perceived as tedious, stressful and
		frustrating, such that surveyed students and professionals do
		not want to write tests in their work or follow the career
		path of a software tester~\cite{DBLP:journals/corr/WaychalC16,
			DBLP:journals/software/WeyukerOBP00}. To motivate developers to test, they
		need variety in their work, creative tasks, recognition, and
		to gain new knowledge \cite{DBLP:conf/esem/SantosMCSCS17}.
		Unfortunately, there is a lack of applicable tools and
		techniques to provide these motivational aspects.

	\subsection{Gamification of Software Testing}
	
	Gamification aims to provide motivation for tasks that are
	normally not one's favorite job, and is commonly interpreted
	as the use of game design elements in non-game
	contexts~\cite{DBLP:conf/mindtrek/DeterdingDKN11}.
	Gamification can lead to more motivation as well as higher and
	better output of the work done by those working in gamified
	environments~\cite{DBLP:journals/ese/StolSG22,
		DBLP:journals/jss/PortoJFF21}.

	The most commonly used gamification elements are points, badges,
	leaderboards, and awards~\cite{DBLP:conf/icse/BarretoF21,
		DBLP:conf/sast/JesusFPF18}. Many more elements can be
	used to gamify activities, depending on the context in which
	gamification is applied. Some elements may subsume other game
	design elements; for example, achievements can be a
	mixture of the game elements points, badges, levels, and awards,
	depending on how they are used or implemented~\cite{robson2015all}.
	An achievement can be gained by making progress in doing a specific task. If the task has been performed a certain number of times, a new level is reached and awarded with points or badges. To get to the next level, the task has to be performed more often than to reach the level before. This can be repeated multiple times until the last level is reached.
	
	Gamification has been demonstrated to improve
		student motivation in learning software
		testing~\cite{DBLP:conf/sbqs/JesusPFS19,DBLP:conf/icsob/Yordanova19}.
		There are several gamified learning
		environments~\cite{fu2016gamification,DBLP:conf/sigcse/FraserGKR19,bell2011secret,elbaum2007bug},
		mostly focusing on unit testing, utilizing elements such as
		points, leaderboards, achievements, and
		levels~\cite{DBLP:conf/sast/JesusFPF18,
			DBLP:conf/profes/MantylaS16}.
	There have also been attempts to gamify aspects of
		testing for professional developers, such as test-to-code
		traceability~\cite{DBLP:conf/sera/Parizi16}, acceptance
		testing~\cite{DBLP:conf/icse/ScherrEH18}, and unit
		testing~\cite{DBLP:conf/icse/StraubingerF22}. The latter
		approach is particularly relevant as our objective is
		similar. However, prior work included gamification as part of
		continuous integration
		(CI)~\cite{DBLP:conf/icse/StraubingerF22}, where
		gamification elements depend on CI builds triggered by
		changes to the version control system. To the best of our
		knowledge~\cite{Fulcini2023}, there have been no previous
		attempts to gamify software testing directly in the IDE
		during development.
	
	\subsection{IDE Support for Testing}
	
	Integrated Development Environments (IDEs) are a crucial part of every
	developer's daily work because they make it easier for them to write
	and test code. A code editor, compiler, debugger, UI builder and
	other tools are
	packed together into one single application with the ability to be
	extended by plugins so that the IDE can be modified to meet any
	requirements~\cite{1463097}.
	Most IDEs support the writing and execution of tests using
	different testing frameworks, which provide an easy
	way to write tests for a specific programming language,
	including execution engines and powerful
	debuggers. In addition, there are many tools available which
	support testing and quality of code in general, like test
	generation \cite{DBLP:conf/icst/ArcuriCF16}, code coverage
	\cite{DBLP:journals/csur/ZhuHM97}, mutation testing
	\cite{5487526} or detecting test smells
	\cite{van2001refactoring}. Consequently, modern IDEs provide
	all the ingredients necessary for effective testing,
	yet so far they are lacking incentives for developers \mbox{to do so.}

%% file: sections/plugin.tex
	\def\HS{\hspace{\fontdimen2\font}}
	
	\begin{table*}[t!]
		\footnotesize
		\centering
		\caption{Overview of implemented achievements with their descriptions and level boundaries}
		%\vspace{-1em}
		\label{tab:codes}
		\begin{tabularx}{\linewidth}{@{}>{}l@{\hspace{.5em}}p{6cm}@{\hspace{.5em}}XXXX}
			\toprule
			\textbf{Achievement} & \textbf{Description} & \textbf{Bronze} & \textbf{Silver} & \textbf{Gold} & \textbf{Platinum}   \\ \midrule
			\textbf{Testing} & & & & & \\ %\midrule
			Test Executor & Execute tests & 3 & 100 & 1,000 & 10,000 \\ 
			The Tester & Run test suites & 3 & 100 & 1,000 & 10,000 \\ 
			The Tester --- Advanced & Run test suites X times containing at least Y tests & X: 10\HS\HS\HS Y: 100 & X: 50\HS\HS\HS Y: 500 & X: 100\HS\HS\HS Y: 1000 & X: 250\HS\HS\HS Y: 3,000 \\
			Assert and Tested & Trigger AssertionErrors & 3 & 10 & 100 & 1,000 \\
			Bug Finder & Previously failed test passes again after source code change & 3 & 10 & 100 & 1,000 \\
			Test Fixer & Previously failed test passes again after test code change & 3 & 10 & 100 & 1,000 \\
			Safety First & Write tests & 10 & 100 & 1,000 & 10,000 \\ \midrule
			\textbf{Coverage} & & & & & \\ 
			Gotta Catch ’Em All & Run test suites with coverage & 3 & 10 & 100 & 1,000 \\
			Line-by-line & Cover lines with your tests & 100 & 1,000 & 10,000 & 100,000 \\
			Check your methods & Cover methods with your tests & 10 & 100 & 1,000 & 10,000 \\
			Check your classes & Cover classes with your tests & 10 & 100 & 1,000 & 10,000 \\
			Check your branches & Cover branches with your tests & 10 & 100 & 1,000 & 10,000 \\
			Class Reviewer - Lines & Cover X classes with at least Y lines by Z\% coverage 
			& X: 5\HS\HS\HS Y: 5\HS\HS\HS Z: 70 
			& X: 20 Y: 25 Z: 80 
			& X: 75\HS\HS Y: 250 Z: 85 
			& X: 250 Y: 500 Z: 90 \\
			Class Reviewer - Methods & Cover X classes with at least Y methods by Z\% coverage 
			& X: 10 Y: 3\HS\HS\HS Z: 60 
			& X: 50 Y: 8\HS\HS\HS Z: 80 
			& X: 250 Y: 15\HS\HS\HS Z: 85 
			& X: 500 Y: 25\HS\HS\HS Z: 90 \\
			Class Reviewer - Branches & Cover X classes with at least Y branches by Z\% coverage 
			& X: 5\HS\HS\HS Y: 15 Z: 75 
			& X: 20 Y: 50 Z: 80 
			& X: 75\HS\HS Y: 250 Z: 85 
			& X: 250 Y: 500 Z: 90 \\ \midrule
			\textbf{Debugging} & & & & & \\ 
			The Debugger & Run the code in debug mode & 3 & 10 & 100 & 1,000 \\
			Take some breaks & Set breakpoints & 10 & 100 & 1,000 & 10,000 \\
			Make Your Choice & Set conditional breakpoints & 3 & 10 & 100 & 1,000 \\
			On the Watch & Set field watchpoints & 3 & 10 & 100 & 1,000 \\
			Break the Line & Set line breakpoints & 3 & 10 & 100 & 1,000 \\
			Break the Method & Set method breakpoints & 3 & 10 & 100 & 1,000 \\
			Console is the new Debug Mode & Use System.out.println instead of debugger or logger & 3 & 10 & 100 & 1,000 \\  \midrule
			\textbf{Test Refactoring} & & & & & \\
			Shine in new splendor & Change source code between two ensuing passing test runs & 5 & 50 & 500 & 2,500 \\
			The Eponym & Rename test method names & 10 & 100 & 1,000 & 10,000 \\
			The Method Extractor & Extract code from tests into a separate method & 10 & 100 & 1,000 & 10,000 \\
			The Method Inliner & Inline methods into tests & 10 & 100 & 1,000 & 10,000 \\
			Double check & Add new assertions to already passing tests & 3 & 10 & 100 & 1,000 \\
			\bottomrule
		\end{tabularx}%
	\end{table*}

	Our approach focuses on integrating gamification elements directly into IDEs, and this is based on several design decisions:

        \begin{itemize}
	\item \textbf{Design Decision 1:} Our goal is to encourage developers to write and use tests. As intrinsic motivation cannot be influenced by external factors, we target extrinsic motivation: Developers should experience satisfaction and a boost in motivation for testing, for which we use \emph{gamification}.
	
	\item \textbf{Design Decision 2:} We recognize that many test interactions occur locally on developers' machines, and not just following CI builds~\cite{DBLP:conf/icse/StraubingerF22}. We therefore aim to motivate \emph{directly in IDEs}. Specifically, we developed a plugin for IntelliJ IDEA from JetBrains\footnote{\url{https://www.jetbrains.com/idea/}}, which is the most widely used IDE in the JVM community~\cite{snyk}. However, the approach itself is not dependent on the specific IDE used.
	
	\item \textbf{Design Decision 3:} Rather than suggesting specific tasks that may disrupt their current workflow~\cite{DBLP:conf/icse/StraubingerF22}, we want to allow developers to choose the testing tasks they find relevant. In addition, we believe that feedback and rewards should be provided quickly, visibly, and immediately after an action to incentivize developers and give them clear goals. We therefore choose the gamification element of \emph{achievements} to provide encouraging feedback immediately when we detect that what a developer did is commendable.
	%The interactions of developers are analyzed in real-time, and rewards are given directly.
	
	\item \textbf{Design Decision 4:} We target short-term interventions, for example, to onboard new developers, or to influence the testing behavior of established developers but nevertheless need to ensure users remain engaged with the gamification as long as they make progress. We therefore provide not only variation in terms of simpler as well as more challenging goals for the achievements, but also different \emph{levels} of achievements and \emph{progress} in between levels.
	
	\item \textbf{Design Decision 5:} We want to nudge developers to participate in testing if they do not do this on their own, and therefore use \emph{notifications} to inform developers of their progress towards available achievements and provide a visual representation of the achievements where the progress is shown.
	
	%    We have chosen to focus on achievements as our gamification element to avoid distracting developers from their workflow.
	
	%    These achievements offer a range of goals and difficulty levels based on best practices in software testing to incentivize developers.
	\end{itemize}

	To implement \toolname with respect to these design decisions, we roughly followed an iterative design research approach: Starting with our own observations, a literature review on gamification of testing as well as developer testing, and a technical inspection of data accessible through IntelliJ's APIs, we defined an initial set of test achievements and implemented a first prototype of \toolname. Feedback on this prototype and its achievements was collected from local researchers and students, and informed refinements as well as further achievements.  We then proceeded to define the study task (\cref{sec:experimenttask}) and procedure (\cref{sec:procedure}), and conducted a pilot study, which informed further improvements, for example by adjusting parameters and thresholds of achievements. After a final round of refinements, we arrived at the version of \toolname also used in the evaluation study presented in this paper.

        \subsection{Testing Achievements}

	When designing the achievements, we tried to adhere to the general criteria~\cite{DBLP:conf/esem/SantosMCSCS17} that good achievements (1) set clear goals, (2)~provide visible progress, (3) offer a variety of tasks, (4) provide recognition, and (5) allow developers to gain new knowledge.
	Each test achievement therefore consists of five elements:
	\begin{enumerate}
		\item A catchy, memorable title.
		\item A description of the aspect that is rewarded.
		\item A progress value that represents the current status %of the user
		with respect to the aspect that the achievement captures.
		\item An underlying mechanism that monitors user actions and updates
		the progress value accordingly. 
		\item Level boundaries, which define when users
		are awarded new levels. For each achievement,
		there are bronze, silver, gold, and platinum levels. For
		example, a developer needs to execute three tests to get the bronze
		level of \textit{Test Executor}, 100 for silver, 1,000 for gold
		and 10,000 for platinum. 
	\end{enumerate}

	The monitoring mechanism for an achievement subscribes to user events
	triggered in the IDE.
	The simplest type of achievement simply counts
	the number of occurrences of an individual type of action, such as how
	often a user executed a test suite, how often assertions were
	triggered, or how often breakpoints were set.
	The progress value can also reflect variables dependent on occurring
	events, such as the number of individual tests executed when the user
	runs a test suite, or coverage levels achieved during such executions.
	Achievements define level boundaries on doing tasks that
	allows progress; for example, progress may be achieved if a certain
	coverage threshold has been reached. % on a certain number of classes.
	Finally, achievements may also monitor not just individual events, but
	more complex sequences of actions that together witness a certain
	testing-related behavior. For example, repairing a broken test
	consists of multiple actions: First the user needs to run a test suite
	in which a test fails, then the test code is edited, and at last the test
	is re-run successfully, all while the code itself remains unchanged.

                We derived objectives that represent good testing behavior based on common testing curricula~\cite{DBLP:conf/sigcse/Jones01, 7328640} and best practices~\cite{DBLP:conf/icse/Kochhar0019, DBLP:conf/sigsoft/Spadini18}, such as:
		(1) ensuring that tests are written and executed after changes, in order to identify bugs in both test and production code~\cite{DBLP:conf/sigsoft/Spadini18, 7328640};
		(2) using metrics such as code coverage to assess the quality of the test suite and identify test gaps~\cite{7328640, DBLP:conf/icse/Kochhar0019};
		(3) using tests to support debugging~\cite{7328640};
		(4) constantly refactoring, adjusting, and simplifying both test and source code to ensure high maintainability~\cite{DBLP:conf/icse/Kochhar0019}.
		Finally, the design of achievements is also constrained by the information about the developer actions that can be extracted from IntelliJ (\cref{sec:implementation}).
                Consequently, we
                %
                %
                % From a user point of view, we
                distinguish between types of achievements
	based on the aspect of testing they refer to:
	%We distinguish categories based on the typical workflow of a developer
	%during testing:
	%
	\begin{itemize}
		\item \textbf{Testing Achievements}: Developers should write and run
		tests, and this category aims to incentivize developers to
		do this more often. Achievements also cover aspects of
		interacting properly with the tests, such as triggering assertions,
		fixing failing tests, or fixing bugs detected by the tests.
		\item \textbf{Coverage Achievements}: %Code coverage is a common
		% approach to assess test suite quality.
		Achievements in this
		category aim to incentivize developers to evaluate 
		the quality of their test suite by using coverage information more
		often, and by improving the coverage achieved by their tests.
		\item \textbf{Debugging Achievements}: If the tests reveal problems,
		they should be used as well as possible to help
		identify the cause. Achievements in this category aim to incentivize the use of
		the IDE's debugging features in conjunction with tests, e.g.,
		running tests in debug mode or with breakpoints set. This implicitly also 
		rewards writing tests, because having good tests is a prerequisite for effective debugging.
		\item \textbf{Test Refactoring Achievements}: %Developers should always strive to improve and refactor their test
		% suites to make them more efficient or readable.
		Achievements in this
		category aim to incentivize developers to improve existing test code
		by applying test refactorings, such as extracting redundant code into
		helper methods.
	\end{itemize}

	Based on these four categories, we define 27 achievements,
	summarized in \cref{tab:codes}. The level boundaries for these
	achievements were determined experimentally during a pilot study (\cref{sec:experimenttask}).
	
	\subsection{User Interface}

	\begin{figure}[t]
		\centering
		\includegraphics[width=0.8\linewidth]{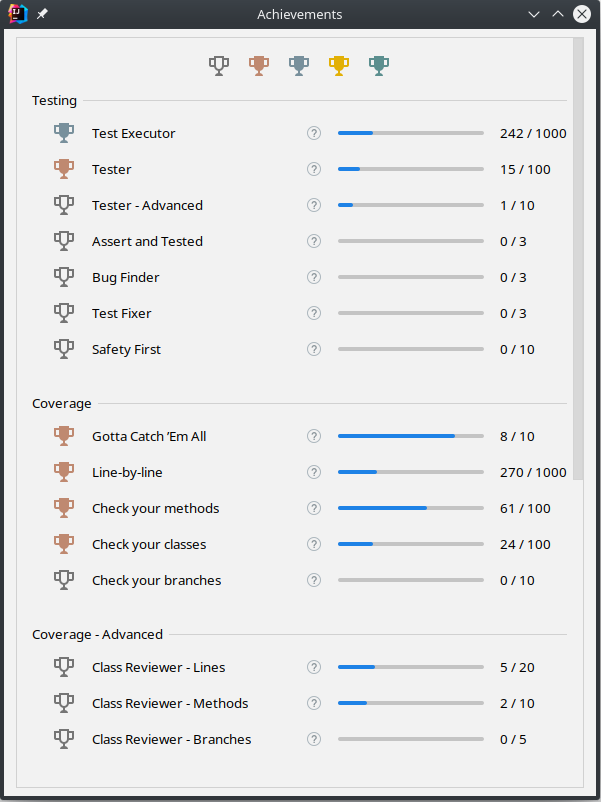}
		%\vspace{-0.5em}
		\caption{IntelliJ window showing achievements and progress}
		\label{fig:achievements}
	\end{figure}

	\begin{figure}[t]
		\centering
		\includegraphics[width=0.9\linewidth]{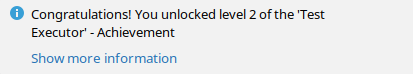}
		%\vspace{-1em}
		\caption{Example notification showing that level 2 of the achievement 'Test Executor' has been reached}
		\label{fig:notification}
	\end{figure}
	
	\begin{figure}[t]
		\centering
		\includegraphics[width=0.9\linewidth]{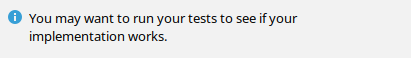}
		%\vspace{-1em}
		\caption{Example notification encouraging to execute tests}
		\label{fig:encouragement}
	\end{figure}

	The achievement overview shown in \cref{fig:achievements}, accessible in the Tools tab of IntelliJ, provides information on available achievements and their current progress. Each achievement is depicted with a trophy, which represents
	the level, a progress bar representing the current progress with
	respect to the next level, and a description of what progress is needed
	to reach the next level, which can be shown by
	hovering over the question mark. Whenever a new level has been reached, the progress bar and the
	information about the next target are updated. The progress value
	itself is not reset after reaching a new level, such that users always
	can see their overall progress. After reaching the final \mbox{(platinum)}
	level for an achievement, the progress value may continue increasing,
	but will not lead to updates of the progress bar because the last
	level has already been reached. At this point, we expect the 
	developers to have adopted that practice and therefore no longer require \toolname.
	The current progress is not project-dependent, new levels can be
	achieved in all opened projects. The progress can also be reset in the
	Help tab in IntelliJ.

	%In addition to the overview window,
	The plugin also shows %three types of
	notifications: First, whenever a new level of an
	achievement has been reached, the user is notified as depicted in
	\cref{fig:notification}; the notification also provides a clickable link to
	the achievement overview. Second, a notification is shown when the developer makes
	progress in one of the achievements. To avoid excessive notifications, these are only shown every 25 \% of progress.  Third, there are notifications
	aimed to incentivize developers to improve progress for
	reaching the next level and solving new achievements (\cref{fig:encouragement}).  Encouraging
	notifications are shown right after installation, and then after a configurable duration without any progress (which we
	set to five minutes for experiments, and 30 minutes in general.)

	\subsection{Implementation} \label{sec:implementation}
	
	IntelliJ is designed to be extensible and was originally implemented
	to support the extension with plugins. There is rich documentation
	about this topic on the website of
	JetBrains\footnote{\url{https://plugins.jetbrains.com/docs/intellij/welcome.html}},
	and IntelliJ itself is open
	source\footnote{\url{https://github.com/JetBrains/intellij-community}}. In
	addition to extending given functionality in IntelliJ, many user
	actions and execution results can be queried, monitored, and analyzed. This is implemented using the
	publisher-subscriber-pattern \cite{DBLP:conf/rtas/RajkumarGS95}, where
	IntelliJ provides several publishers, to which a plugin can subscribe 
	to then be notified about changes or user interactions, such as
		test runs and their results,
		changes in the project's structure and files,
		coverage after a test run with coverage, and 
		information on debugger and breakpoints usage.
	
	To implement achievements of the category Test Refactoring we use RefactoringMiner~\cite{DBLP:journals/tse/TsantalisKD22},
	which compares edited files before and after a change and derives information about applied refactorings. In particular, \emph{Shine in new splendor} is implemented by checking whether RefactoringMiner reports any refactorings for a change (that follows a test execution and is succeeded by another one). We use the \emph{renaming} refactoring to identify \emph{The Eponym}, the \emph{extracting} refactoring for \emph{The Method Extractor}, and the \emph{inlining} refactoring for \emph{The Method Inliner}. \emph{Double check} is implemented by directly comparing successive versions of test code.

%% file: sections/evaluation.tex
	In order to evaluate whether the gamified development environment
	influences the developers' behavior, we conducted a controlled
	experiment and aim to answer the following research questions:
	\begin{itemize}
		\item \textbf{RQ 1}: Does \toolname influence
		 testing behavior?
		\item \textbf{RQ 2}: Does \toolname influence
		 resulting test suites?
		\item \textbf{RQ 3}: Do achievement levels reflect differences in test suites and activities?
		\item \textbf{RQ 4}: Does \toolname influence the functionality of resulting code?
		\item \textbf{RQ 5}: Does \toolname influence
		the developer experience?
	\end{itemize}
	
	\subsection{Experiment Setup}
	
	The controlled experiment, consisting of a programming task, was carried out at the \passau in multiple
	sessions in June and July 2022 as well as February 2023.
	
	\subsubsection{Experiment Task} \label{sec:experimenttask}
	
	To improve ecological validity of the experiment, we aimed to
	base the programming task on code taken from a real software
	project. In addition, the task should be challenging, implementable 
	and testable in 60 minutes, have a good specification for clear and easy understanding
	as well as no complex structures and dependencies.
	We found suitable candidate tasks in the work of Rojas et al.
	\cite{8367053}, since their selected classes from Apache Commons %\footnote{\url{https://commons.apache.org/}} 
	meet these requirements. From this data set, we used the
	\textsf{FixedOrderComparator}
	class\footnote{\url{https://commons.apache.org/proper/commons-collections/apidocs/org/apache/commons/collections4/comparators/FixedOrderComparator.html}},
	because every developer should be familiar with the
	general concept of a comparator, which reduces the time needed for understanding the task. In addition, the class and its
	functionality are easy to understand and test, and Rojas et al. revised
	and extended the JavaDoc specification.
	
	We conducted a pilot
	study with five participants, the data of which is not
	included in the final experiment and our analysis. The participants of the
	pilot study were required to implement and test the whole
	Java class using JUnit tests from scratch, half of them with and without \toolname. After the study, it became clear that 60
	minutes would not be sufficient time to complete the
	implementation and write adequate tests. Therefore, we
	simplified the task by providing a scaffolding consisting of
	the class with the contents of the two methods
	\textsf{addAsEquals()} and \textsf{compare()} removed except
	for the signature and JavaDoc comments.
	
	Furthermore, we used the data gathered during the pilot study to 
	adjust the level boundaries of the achievements to values realistically usable in practice: We extracted the numbers 
	of actions performed by the participants and set the level boundaries so that 
	it would take a few weeks to reach the platinum level when working continuously.
	Not every boundary could be extracted during the pilot study, because the participants did not work on every achievement, and we estimated appropriate values for the remaining ones.
	%,
	%since there are other possibilities to ensure the functionality like
	%manual testing.
	
	In addition to the \textsf{FixedOrderComparator} class, a \textsf{Main}
	class was given to the participants, which makes it easier to apply
	manual testing, i.e., testing within the main method. We also prepared the
	folder structure including a test folder, such that there would be no
	configuration effort for participants to add tests, if they decided to add a test class	(e.g., \textsf{FixedOrderComparatorTest}).

	\subsubsection{Participant Selection}
	
	To recruit participants we created an eligibility survey
	consisting of basic demographic questions, questions about programming
	experience in Java and different testing tools for Java, as
	well as five technical questions about JUnit to assess the
	testing knowledge. % in Java with JUnit.
	Each of the single-choice technical questions is based on a
	small code example with four answer options\footnote{\label{not:artifacts}
		Detailed information included in the artifacts}. We advertised the survey to all (former)
	computer science students who had taken one of our software
	engineering-related courses within the last year. Thus, all
	participants had previously passed courses with advanced
	Java-based programming assignments and mandatory
	coverage-based unit testing, which increases confidence that
	they are familiar with Java, JUnit, and debugger and
	coverage tools.
	
	We received 62 responses to the survey. As minimum
	qualification we required at least three out of the five
	technical questions answered correctly as evidence of
	expertise, which 49 respondents satisfied and were therefore
	chosen as our participants; 86 \% of the participants were
	students, while the rest were graduates now working in
	industry (8~\%) or university (6~\%). The age ranges from 20
	to 34, with 43 male and six female participants. Most
	participants (76 \%) claimed to have more than three years of
	experience with Java, and more than one year with JUnit (59
	\%).
	Each participant received a 15~€ Amazon voucher as an expense
	allowance.

	\subsubsection{Experiment Procedure} \label{sec:procedure}
	
	The experiment was conducted in multiple sessions %based on the
	% participants' availability
	in person in the computer lab of the \passau.
	Each experiment session started with a 15-minute introduction
	explaining the experiment procedure and the
		programming task based on examples and documentation of how the class should behave. We furthermore provided each participant with a
	printed specification of the two methods to implement. 
	
	We  divided participants into four groups:
	\begin{itemize}
		\item \textbf{Control group}: Participants have to solve the task without the plugin.
		\item \textbf{Treatment group}: Participants have to solve the task with the plugin enabled, without being required to use it.
		\item \textbf{Notifications group}: Participants have to solve the task with motivational notifications shown during the development process but without visible achievements.
		\item \textbf{Maximizing group}: Participants have to solve the task with the plugin, and they are explicitly asked to solve as many achievements as possible.
	\end{itemize}

	The \control and \treatment groups are our main analysis focus since comparing these two groups reveals the difference in developing with and without gamification.
	Since the plugin shows notifications together with achievements, it is unclear whether
	any observed effects result from the achievements or are simply triggered by notifications. To discern these effects, we
	evaluate the notifications as a stand-alone feature with the \notifications
	group, which receives only generic notifications (e.g., \cref{fig:encouragement}) but no notifications about achievements.
	A further concern may be whether any observed differences are a result of participants focusing on the achievements rather than programming, for example, because they incorrectly assume that is what we expect them to do. To distinguish such effects from participants focusing on development, the \maximizing group consists of participants explicitly tasked to solve as many achievements as possible.
	
	We organized six experiment sessions in two blocks, wherein the first one (4 sessions) the participants were assigned to either the \control or \treatment groups, and in the second block (2 sessions) to the \notifications or \maximizing groups. The participants were alternatingly assigned to the groups based on the sequence in which they entered the room. We assigned more participants to the \control and \treatment groups since these represent our main focus.
	Of the 49 participants, 17 were assigned to the \treatment group, 
	16 to the \control group and eight each to the \notifications and \maximizing groups. 
	In addition to general explanations\footnoteref{not:artifacts}, 
	participants in the \treatment and \maximizing groups were shown the plugin individually.
        The goal of the study was not revealed until after the study to avoid biasing behavior. 
	
	Following the introductory explanations, participants had 60
	minutes to implement the Java programming task in an IntelliJ
	environment. % prepared by us.
	The task was (1) to implement the methods \mbox{\textsf{addAsEquals()}}
		and \textsf{compare()} based on the explanations and documentation given\footnoteref{not:artifacts}, and (2) to ensure the correctness of the methods' functionality.
	We
	did not explicitly instruct participants to write unit tests,
	but discussed possibilities like unit tests or manual testing
	with the main method.
	Throughout the experiment, a custom event logger saved the
	states of the achievements after each user interaction.
	%in a
	%CSV file.
	A
		custom script committed and pushed the current implementation
		and log files to a Git repository once per minute. % to the respective
	%branch. 
	
	% From the main branch, every participant got their own
	%branch checked out where they work on.

	%The
	%logs were stored in the resources directory, with an additional log
	%only showing the test executions.
	
	After finishing the 60-minute implementation phase,
	participants completed an exit survey online, consisting of
	general questions about the implementation and testing of the
	class. A second page, shown only to the \treatment and
	\maximizing groups, asked about experiences with the
	plugin. Answer options were based on a five-point Likert
	scale, and the questions can be viewed in
	\cref{sec:rq4exit}. An optional free-text question allowed
	to provide further comments.

	\subsubsection{Experiment Analysis}
	
	The analysis of the experiment is based on comparing results between
	the four groups. To determine
	the significance for any of the measurements between those groups, the
	exact Wilcoxon-Mann-Whitney test \cite{10.1214/aoms/1177730491} was
	used to calculate the $p$-values with $\alpha = 0.05$. 
	%\todo{Ideally we should also state effect sizes, not only p values. Depends on how much time is left.}
	When visualizing trends we show the 84.6 \% confidence
        intervals for calculating the
        means~\cite{fisher1956statistical}; if the intervals overlap
        there is no statistically significant
        difference~\cite{DBLP:journals/csda/AfshartousP10} (which is
        equivalent to checking significance using an exact
        Wilcoxon-Mann-Whitney test with $\alpha = 0.05$).
	
	%\todo{At some place we should also add a URL with the anonymized data/scripts}
	
	\paragraph{RQ 1: Does \toolname influence testing behavior?}
	
	We compare the testing behavior in terms of (1)~test creation, (2)~test executions, (3)~coverage measurement, and (4)~use of the debug
	mode to run tests.
	Since the programming task involves an API rather than a program with
	a user interface, it is always necessary to write some test driver
	code to exercise the implemented code. The desirable approach to do so
	is to put this code into JUnit tests; a less desirable approach, akin
	to manually testing a program, is to temporarily place this code into
	the \textsf{Main} class. Even though this code is typically deleted
	again afterward, we can consider the entire history of the code and
	thus determine how often manual testing was applied.  We consider all
	methods annotated using the JUnit \textsf{@Test} annotation as tests,
	regardless of their content.  The number of test executions, with and
	without coverage or debugging, is collected by our plugin, 
	which is running even if the UI is not visible (for the \control and \notifications groups).
	We also consider the number of tests, 
		test executions with and without coverage, and debug uses over time based on the commit history. In addition, we calculate significant differences in testing between the groups with the exact Fisher test~\cite{bower2003use} with $\alpha = 0.05$.
	
	%We also consider the number of test executions with and without
	%coverage report in total and over time to answer this research
	%%%question, as well as the total number of usages of the debug mode. Both the total number and the evolution over
	%time via commit history could be determined with our data. The number
	%of test runs with coverage reports could also be determined in the
	%same way, as well as the number of times, the debug mode was used. 
	
	\paragraph{RQ 2: Does \toolname influence resulting test suites?}
	
	To answer this question, we compare the four experimental groups in terms
	of (1) the number of tests, (2) code coverage, and (3) mutation scores
	of the final test suites. We again take only methods with the
	\textsf{@Test} annotation into account as tests.
	We measured the code coverage of the implementation of the
	participants with the help of
	JaCoCo\footnote{\url{https://www.jacoco.org/jacoco/}}.  We decided to
	not use the internal code coverage report of IntelliJ because JaCoCo
	is more widely used and can be automated. Failed tests
	were excluded from coverage measurement.
	%, because most of those tests were
	%written just before the end of the experiment and were therefore not
	%completely finished.
	Only the two methods to be implemented as well as added helper
	methods were included in the coverage measurements. We used both line and branch coverage to
	get a complete overview of the covered code.
	To calculate mutation scores we used the PIT\footnote{\url{https://pitest.org/}} mutation testing system for
	Java. PIT was integrated into each project of the
	participants as a Maven plugin and configured %in the \textsf{pom.xml}
	% file
	to generate mutants for the methods \textsf{addAsEquals()} and \textsf{compare()} only. As constructors could
	not be excluded in the mutation analysis, we manually removed these mutants
	for calculating the mutation scores. Since PIT expects test suites without failing tests, all
	failing tests were excluded. In all cases, these are tests the participants were actively working on but did not have the time to complete when the experiment ended.
	
	\paragraph{RQ 3: Do achievement levels reflect differences in test suites and activities?}
	
	\toolname logs the current progress of each achievement as well as
	the reached levels after each user interaction during the experiment.
	For each participant, we extract the total number of times a
	new level was reached on any of the achievements in the end and during the
		study to extract the levels reached every minute during the study.
	%
	%These levels are summed up for each participant to get
	%the total number of reached achievement levels.
	We then compare this number between the four groups and
	measure the Pearson rank correlation~\cite{pearson1895notes}
	with line coverage, branch coverage, mutation score, and number
	of tests. % to answer this research question.
	
	\paragraph{RQ 4: Does \toolname influence the functionality of resulting code?}
	
	To answer this question, we used the golden test suite from Rojas et
	al.~\cite{8367053}, which consists of six test cases. We executed
	these tests against the final version and all intermediate versions based on the commit history of each participant's code, and
	compare the numbers of passing tests between groups.
	
	\paragraph{RQ 5: Does \toolname influence the developer experience?}
	
	This question is answered by comparing the answers to the exit survey. %conducted at the end of the experiment.
	For better
	visualization, the data is presented in stacked bar charts
	including the questions and percentages.

	\subsubsection{Threats to Validity}
	
	There are potential \emph{threats to external validity} of our study. We only focused on one specific implementation task, which means that our findings may not be applicable to other tasks. However, since unit testing, debugging, and refactoring are fundamental concepts used in various programming languages and applications, we believe that our approach can be generalized to a broader context.
	Additionally, our experiment was conducted with participants from a specific university only, and results may differ if the experiment was repeated with participants from other universities or from industry.
	Furthermore, our evaluation of \toolname was limited to a short-term experiment, which may restrict the generalizability of our findings to long-term usage scenarios. However, \toolname is designed to be adaptable and can be easily extended for long-term studies.

	\emph{Threats to internal validity} may arise from errors in our data
	collection infrastructure, plugin, or experiment procedure. We
	thoroughly tested all software and validated it in a pilot
	study. Furthermore, the groups could be imbalanced with respect to demographics or programming skills. However, there were no significant differences in age, gender, and occupation between the groups according to the exact Wilcoxon-Mann-Whitney test \cite{10.1214/aoms/1177730491} with $\alpha = 0.05$. In terms of expertise, there were no significant differences except that the \treatment group claimed more self-declared expertise in Java ($p=0.011$) and JUnit ($p=0.017$) compared to the \maximizing group, which, however, does not negatively affect any of our conclusions.
	Our qualification
	questions may not be precise enough to fully capture 
	knowledge and experience. To mitigate this potential issue, we selected participants who had been enrolled in a course that covered unit testing. Additionally, we assessed their understanding of testing with JUnit by asking five questions, and out of 245 responses (five questions for each of the 49 participants) we only identified 16 incorrect answers. A perfectly balanced distribution of participants would likely require a closer assessment of skills, which would inhibit recruitment.

        Our recruitment process may have introduced a selection-bias~\cite{kenny1979returns} as all participants voluntarily chose to take part. A higher number of participants who have a natural inclination or motivation towards testing may skew results in terms of more tests, achieved levels, or coverage compared to what an average population might achieve. To mitigate this threat, we only advertised the experiment as programming task without explicitly mentioning the aspect of testing or gamification, thus hoping for a more representative sample of students with interest in programming.
	To avoid that prior knowledge of the
	\mbox{\textsf{FixedOrderComparator}} class influences results or that the
	code would be found through a web search, all references to the
	originating software project and authors were removed. To minimize
	external influences we conducted the
	experiment in a controlled environment at the \mbox{\passau.}
	
	There are potential \emph{threats to construct validity} of our study. One concern is that the golden test suite may not encompass all possible edge cases. The golden test suite achieves 90\,\% line coverage and an 86\,\% mutation score in both methods; while it does not fully cover statements related to input validation, it does test everything related to the main functionality. Another potential threat is the setup of our experiment, which took place in a controlled lab session. Like all controlled environments this may not accurately represent the experiences and behaviors of all developers in real-world scenarios.

	%	\emph{Threats to construct validity} arise by the mere visibility 
	%	of the plugin, which could lead to an increased number of tests without gamification. With our current experimental setup, it is not possible to distinguish the effects between different achievements.

%% file: sections/results.tex
	\subsection{RQ 1: Does \toolname influence testing behavior?}
	
	%To answer this research question, five sub-questions will be analyzed and answered.
	
	\paragraph{Test Creation}
	\toolname intends to incentivize writing unit tests, and
	appears to achieve this: There was only one participant in the
	\treatment group who did not write JUnit tests at all, but six
	participants of the \control group which is significant ($p=0.039$) according to the exact Fisher test~\cite{bower2003use}.
	In contrast, participants of the \control group used the main
	method for manual testing more frequently: Only 2/17
	participants in the \treatment group used the main method for
	testing at some point, compared to 7/16 of the \control group.
	Notifications show some effect, but less than the
	achievements: 2/8 participants wrote no
		unit tests, and 2/8 used main testing which can be seen as an indication that the participants of the \notifications group are slightly more motivated than the \control group, but less
		than the \treatment group.
	Focusing on achievements seems to lead to the most motivation~\cite{8370133} to
	test: All participants of the \maximizing group wrote unit
	tests, and only one of them used main testing, which suggests that 
		achievements make developers test more than needed to complete the study task.

	\begin{figure*}
		\centering
		\begin{subfigure}[t]{0.33\textwidth}
			\centering
			\includegraphics[width=\textwidth]{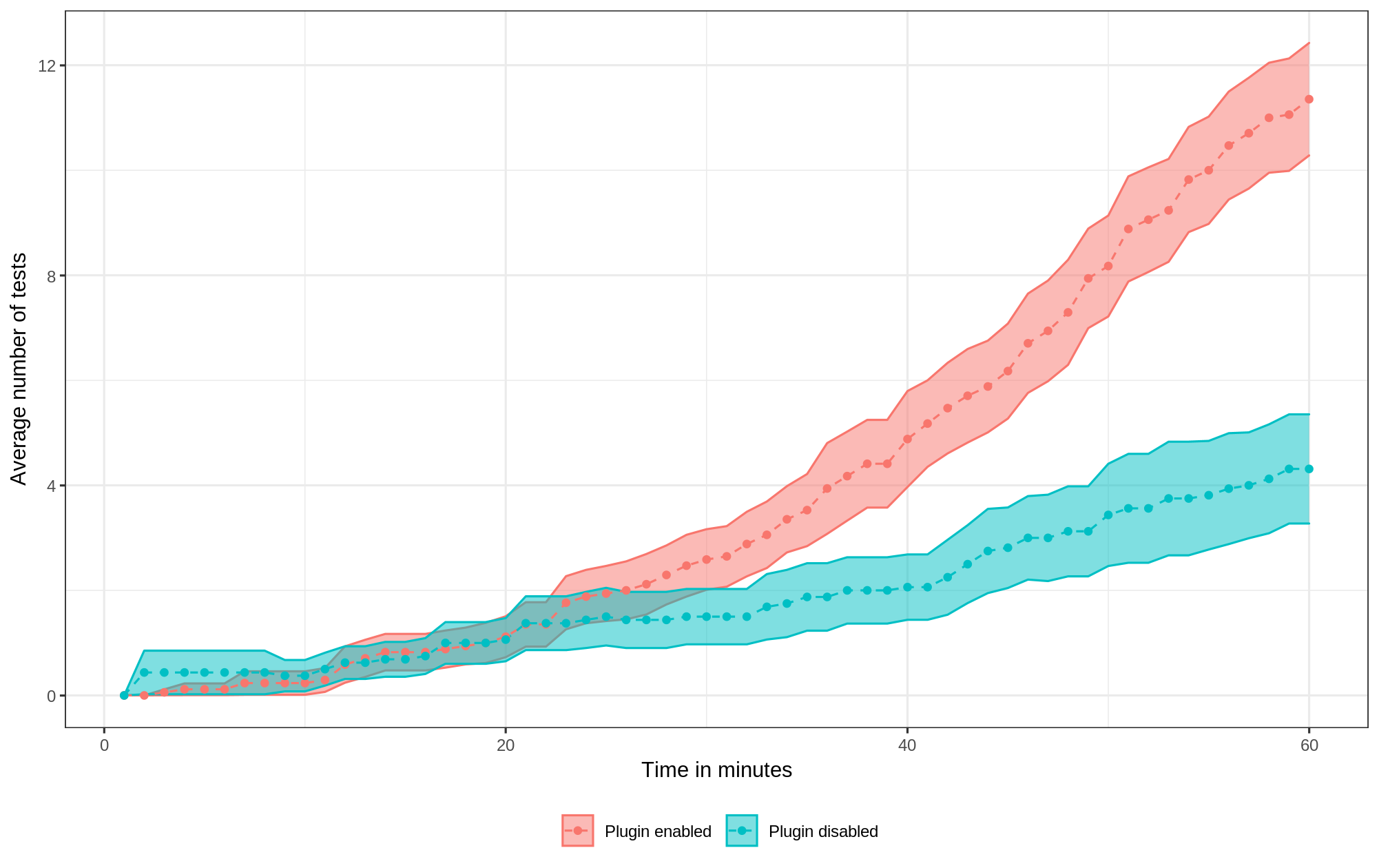}
			\vspace{-1em}
			\caption{Number of tests created over time}
			\label{fig:teststime}
		\end{subfigure}
		\hfill
		\begin{subfigure}[t]{0.33\textwidth}
			\centering
			\includegraphics[width=\textwidth]{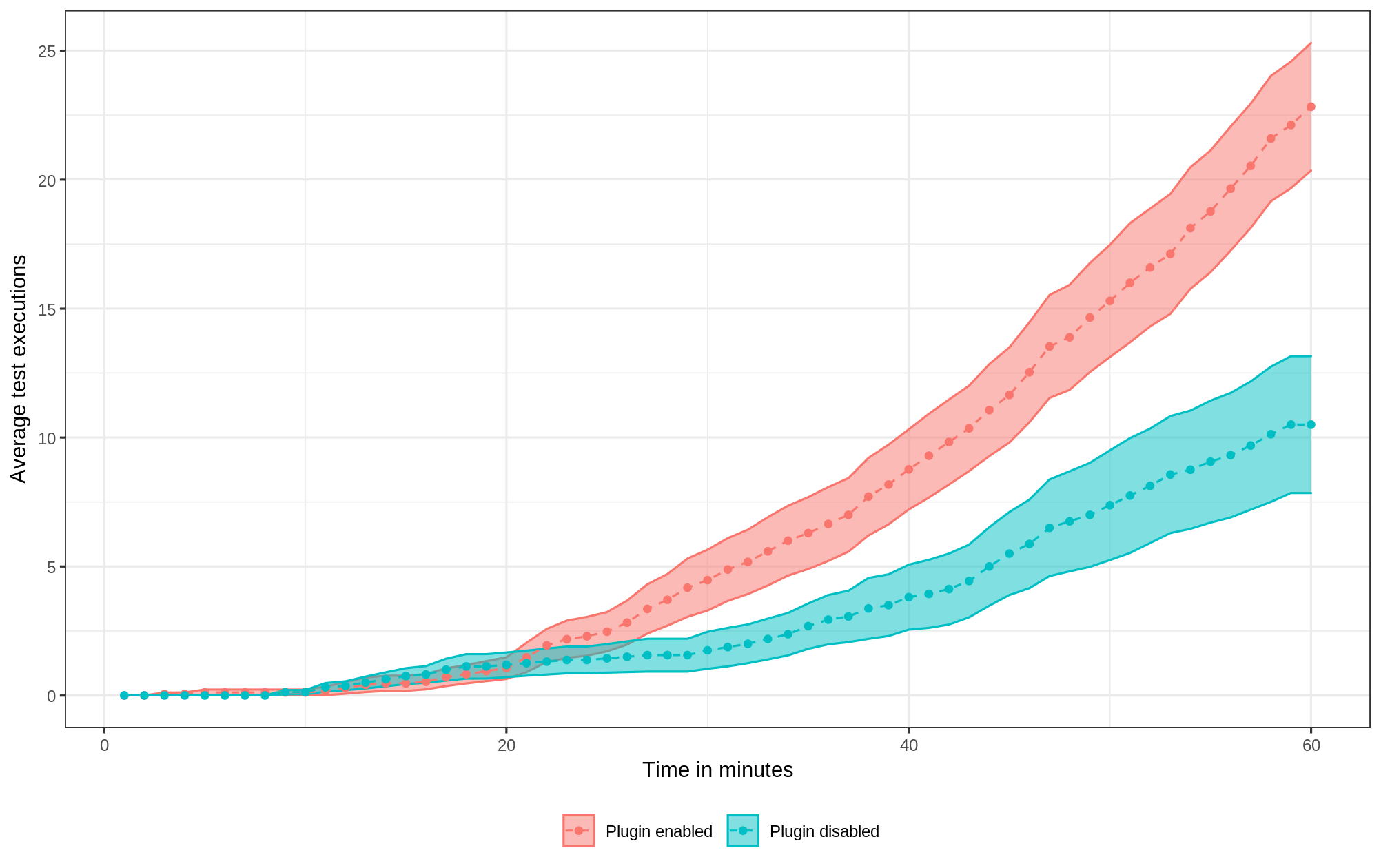}
			\vspace{-1em}
			\caption{Number of test executions over time}
			\label{fig:testexecutionstime}
		\end{subfigure}
		\hfill
		\begin{subfigure}[t]{0.33\textwidth}
			\centering
			\includegraphics[width=\textwidth]{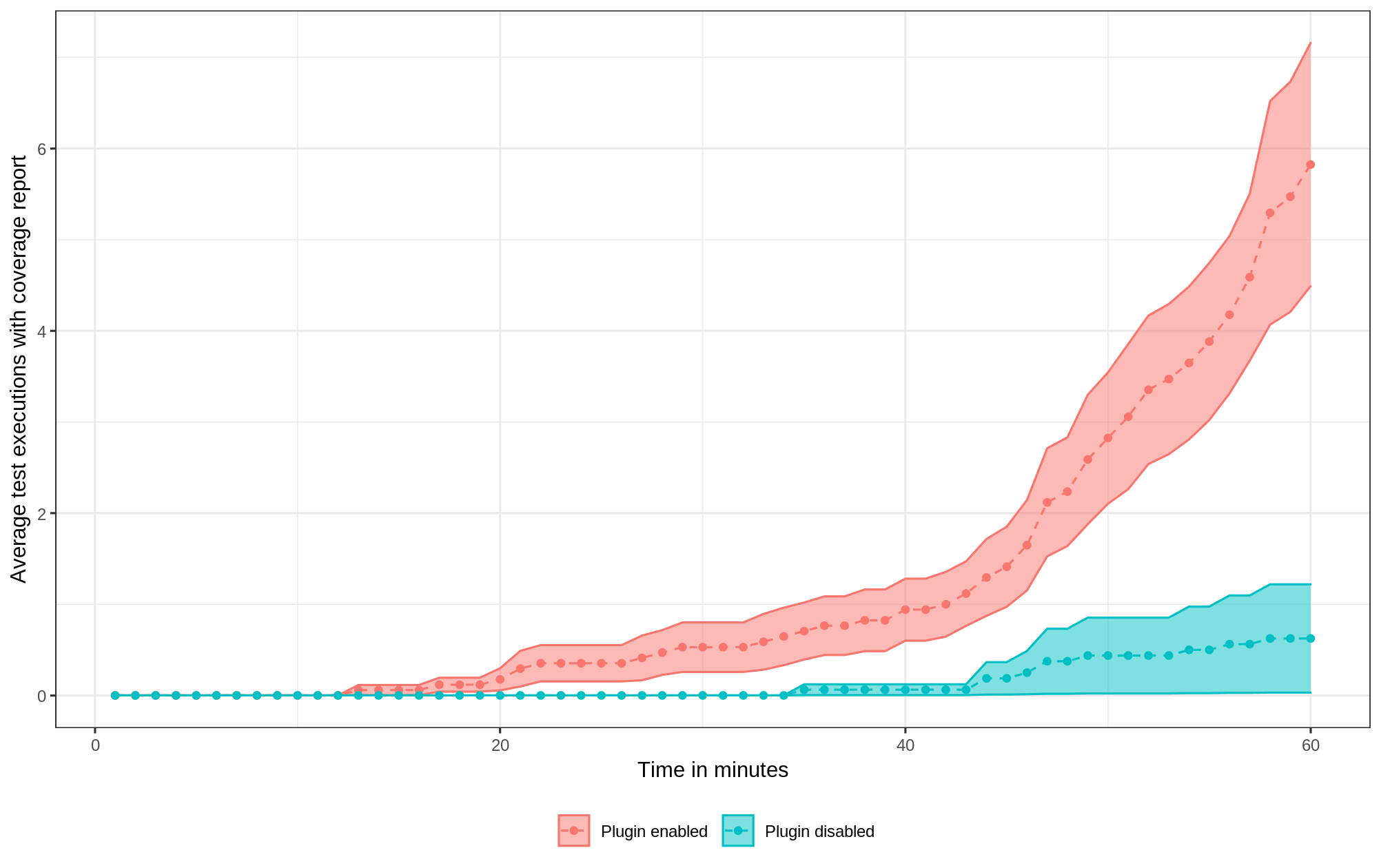}
			\vspace{-1em}
			\caption{Number of test executions with coverage over time}
			\label{fig:testexecutionscoveragetime}
		\end{subfigure}
		\hfill
		\begin{subfigure}[t]{0.33\textwidth}
			\centering
			\includegraphics[width=\textwidth]{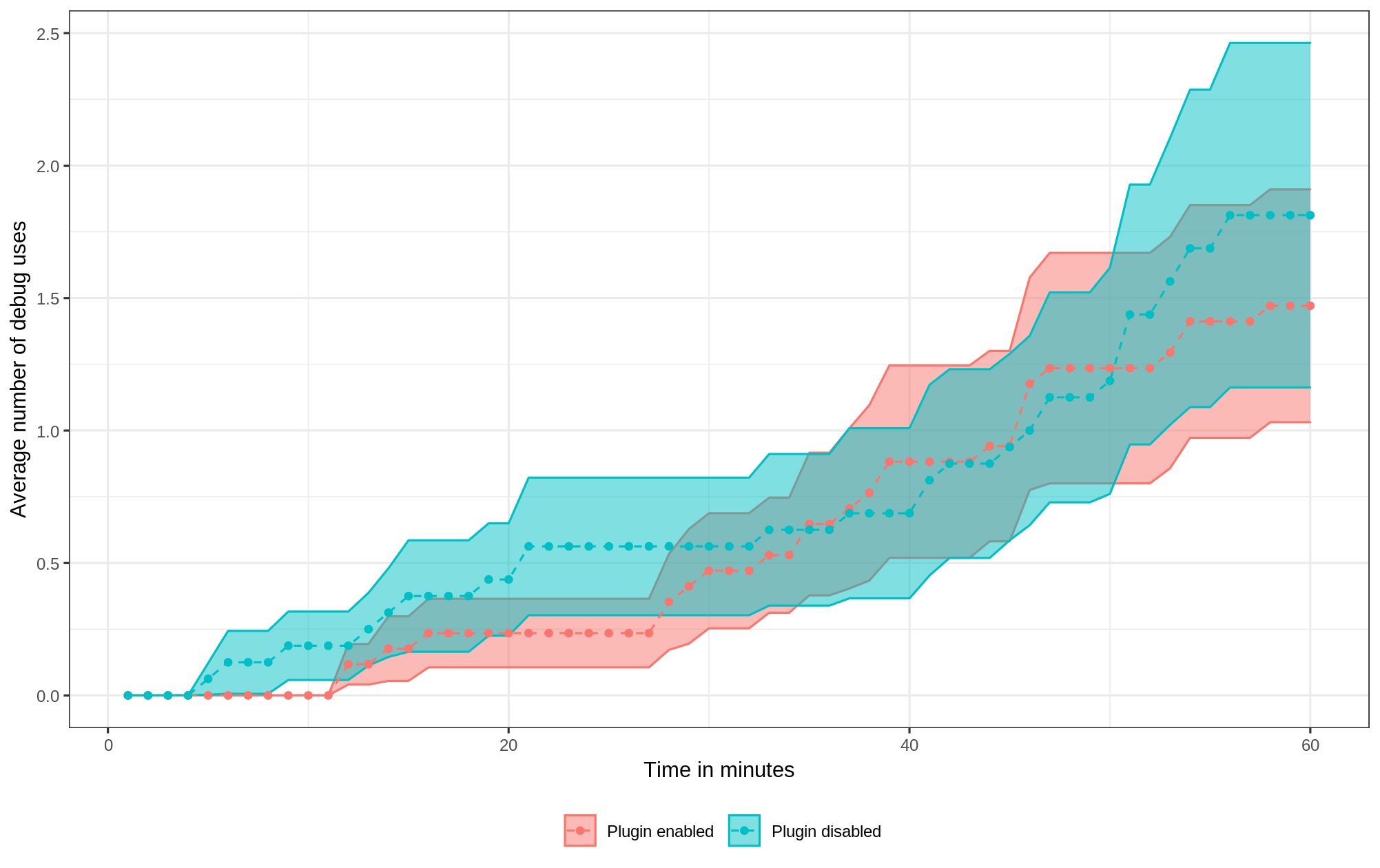}
			\vspace{-1em}
			\caption{Number of debug uses over time}
			\label{fig:debugtime}
		\end{subfigure}
		\hfill
		\begin{subfigure}[t]{0.33\textwidth}
			\centering
			\includegraphics[width=\textwidth]{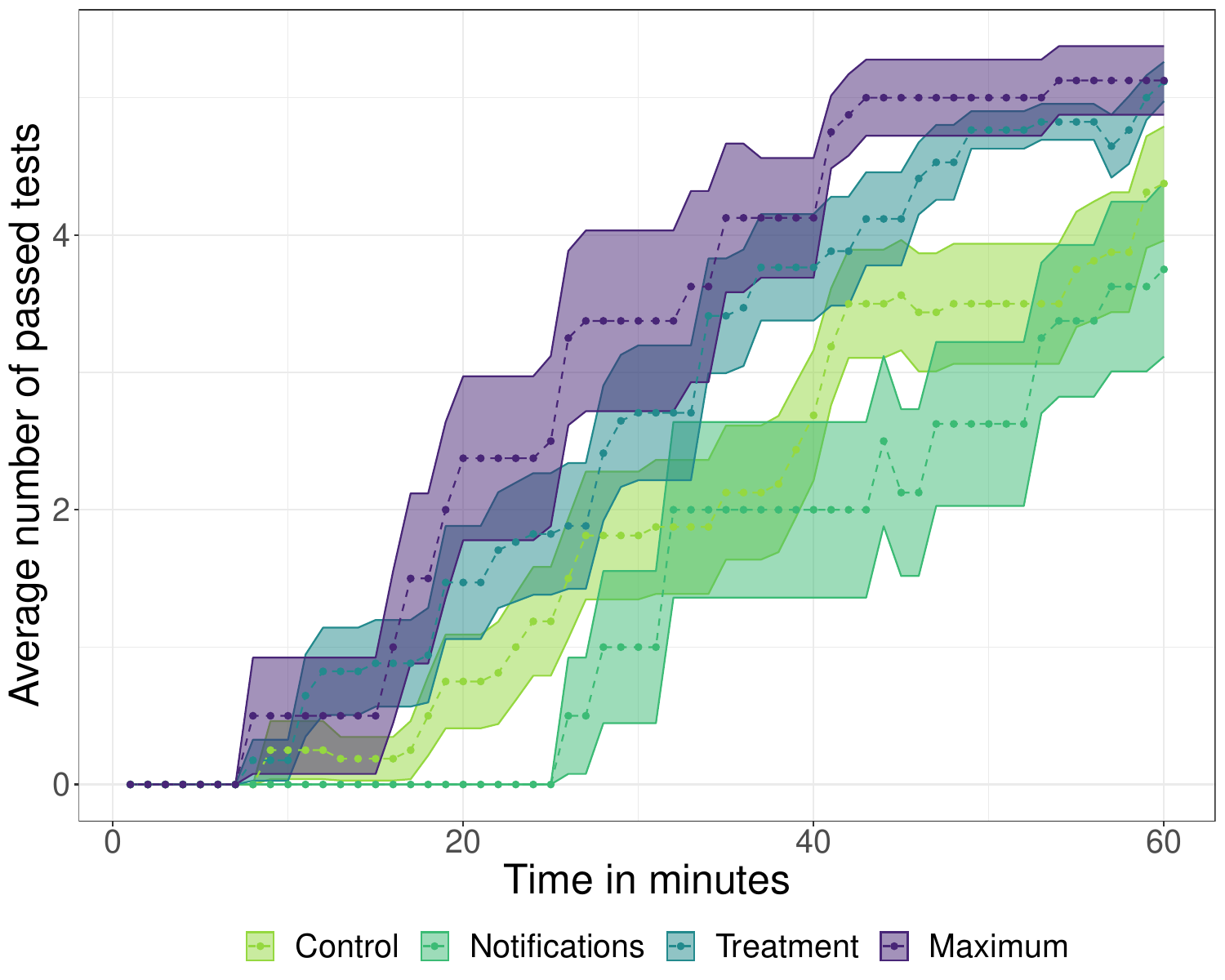}
			\vspace{-1em}
			\caption{Number of passed tests over time}
			\label{fig:passedteststime}
		\end{subfigure}
		\hfill
		\begin{subfigure}[t]{0.33\textwidth}
			\centering
			\includegraphics[width=\textwidth]{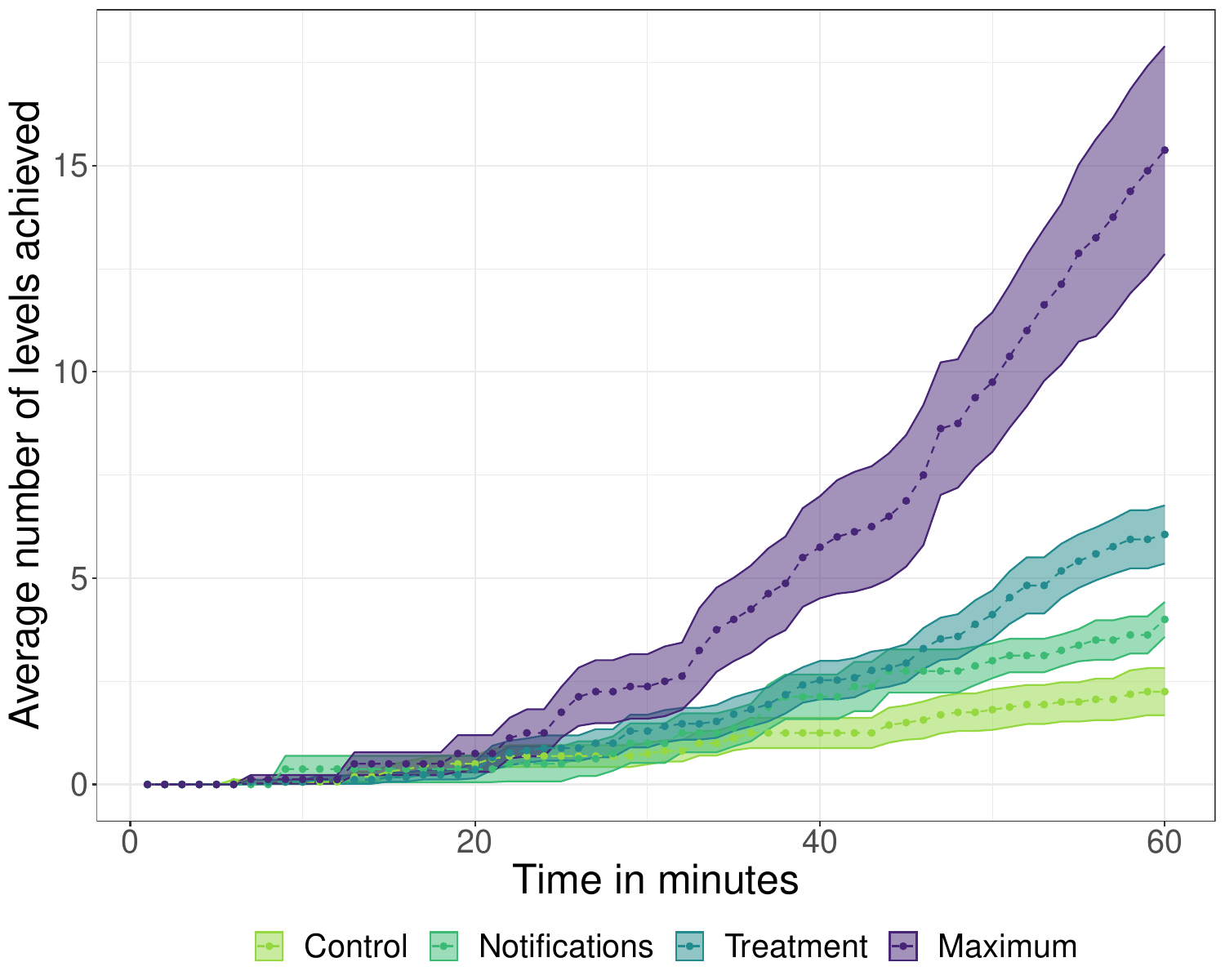}
			\vspace{-1em}
			\caption{Number of levels over time}
			\label{fig:levelstime}
		\end{subfigure}
		
		\caption{Differences between \control, \notifications, \treatment and \maximizing groups over time (see \cref{sec:procedure})}
		\label{fig:timeplots}
	\end{figure*}

	\Cref{fig:teststime} shows the JUnit tests written 
	over time: Both the \treatment and \control groups start early to write tests; the first
	participant of the \control group wrote a test after a minute, and for
	the \treatment group after two minutes. 
	From 22 minutes onwards the number of tests in the \treatment
	group increases much faster than for the other groups (except \maximizing), and from minute 30 on 
	the improvement over the \control group is significant (the
		confidence intervals do not overlap.)
	%Before starting to test, most of the participants implemented parts of the logic we required in the experiment.
	%
	The \notifications group initially behaved similarly to the \treatment group, but stagnated towards the end and stayed between both the \control and \treatment groups, confirming that notifications contribute to, but do not fully explain, the improvements. As expected, the \maximizing group started sooner and wrote significantly more tests than any other group between minutes 31 and 50, at which point the \treatment group caught up. 
	%
	%	Overall, participants of the \treatment and \maximizing groups wrote significantly more JUnit tests in the second half of the
	%	experiment than the other groups.
	
	\paragraph{Test Executions}
	\label{sec:runtests}
	
	\begin{figure*}
		\centering
		\begin{subfigure}[t]{0.245\textwidth}
			\centering
			\includegraphics[width=\textwidth]{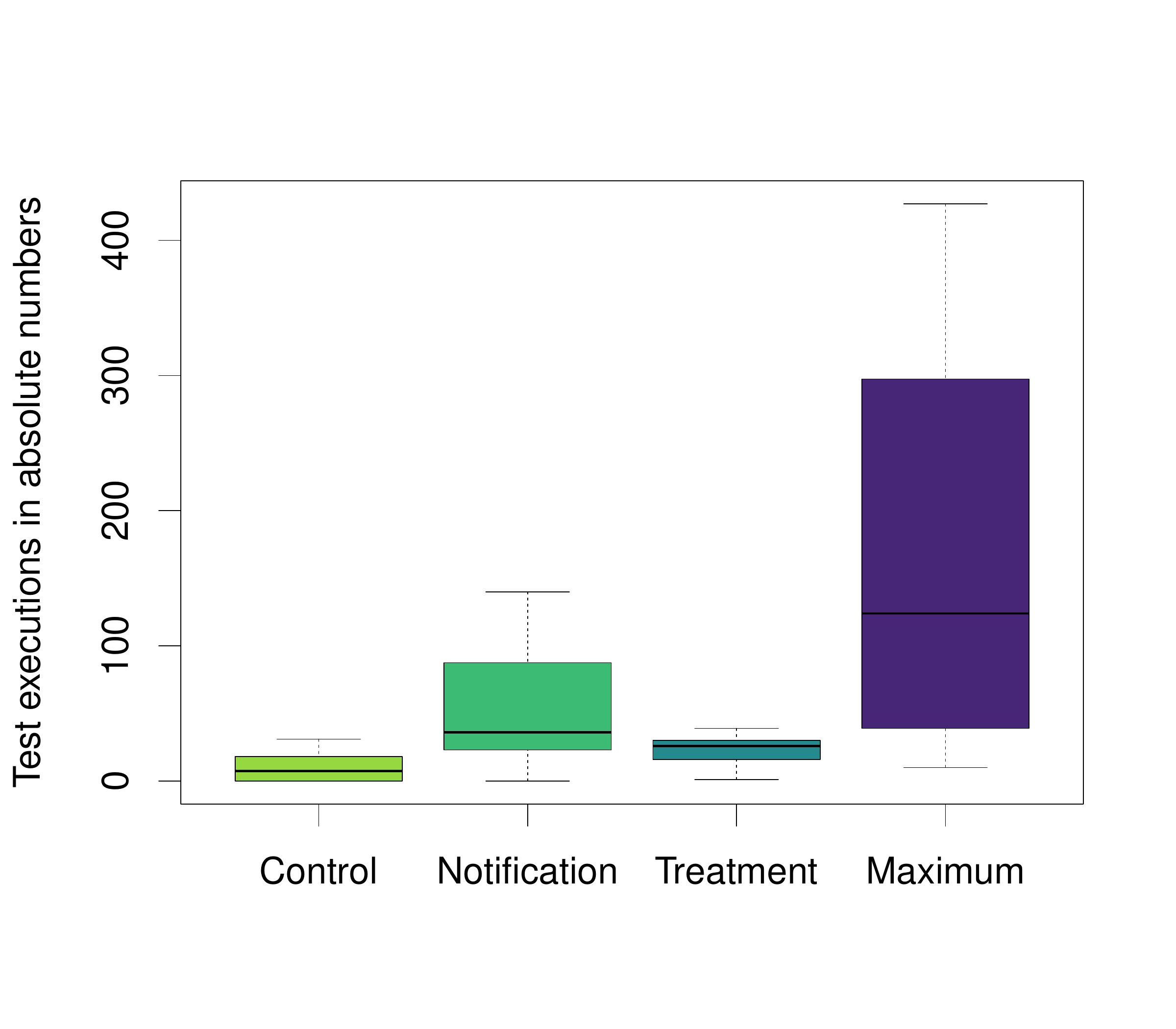}
			\vspace{-2.5em}
			\captionsetup{justification=centering}
			\caption{Number of test executions \\during the experiment}
			\label{fig:testexecutions}
		\end{subfigure}
		\hfill
		\begin{subfigure}[t]{0.245\textwidth}
			\centering
			\includegraphics[width=\textwidth]{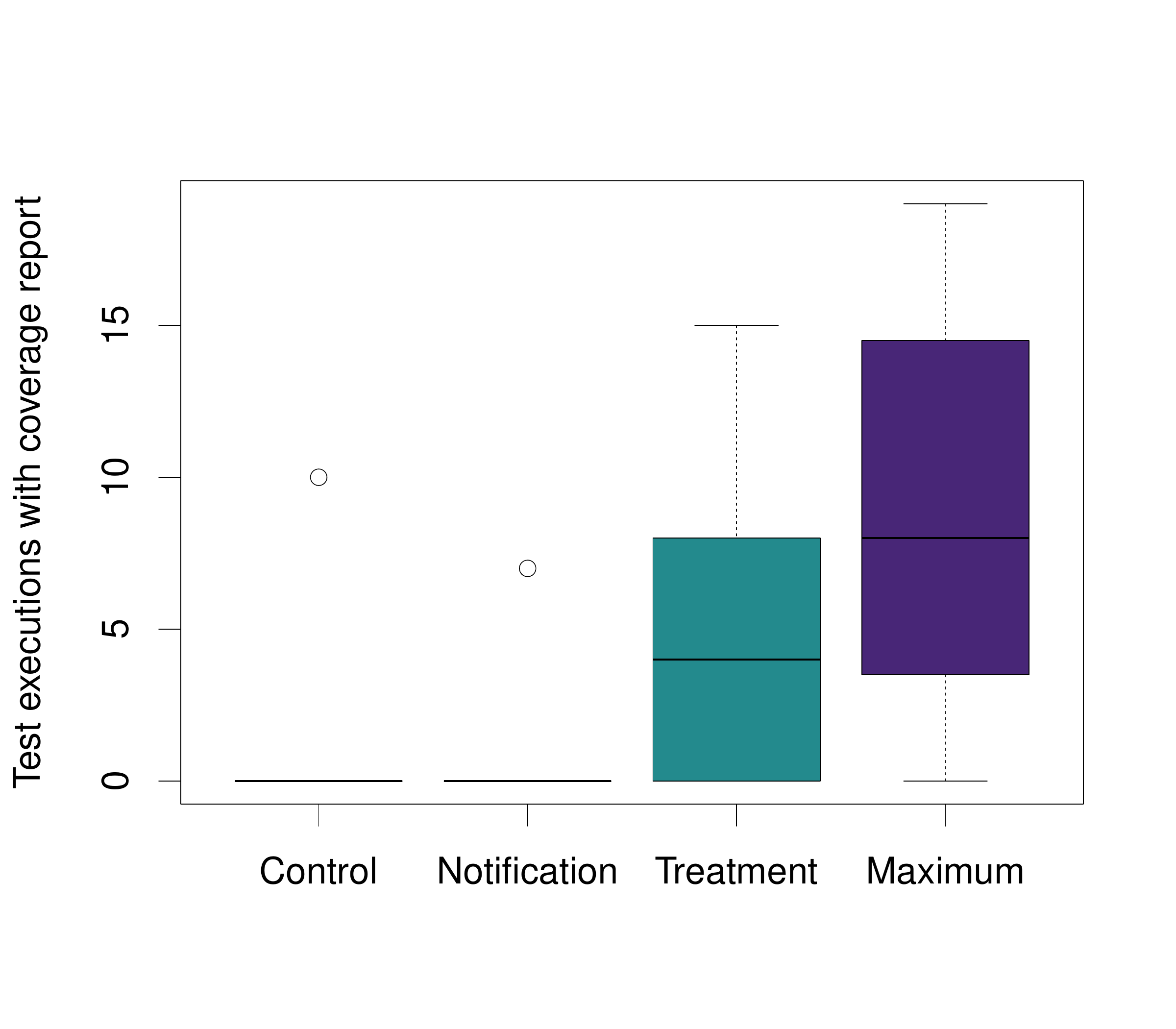}
			\vspace{-2.5em}
			\captionsetup{justification=centering}
			\caption{Number of test executions \\with coverage in IntelliJ}
			\label{fig:testexecutionscoverage}
		\end{subfigure}
		\hfill
		\begin{subfigure}[t]{0.245\textwidth}
			\centering
			\includegraphics[width=\textwidth]{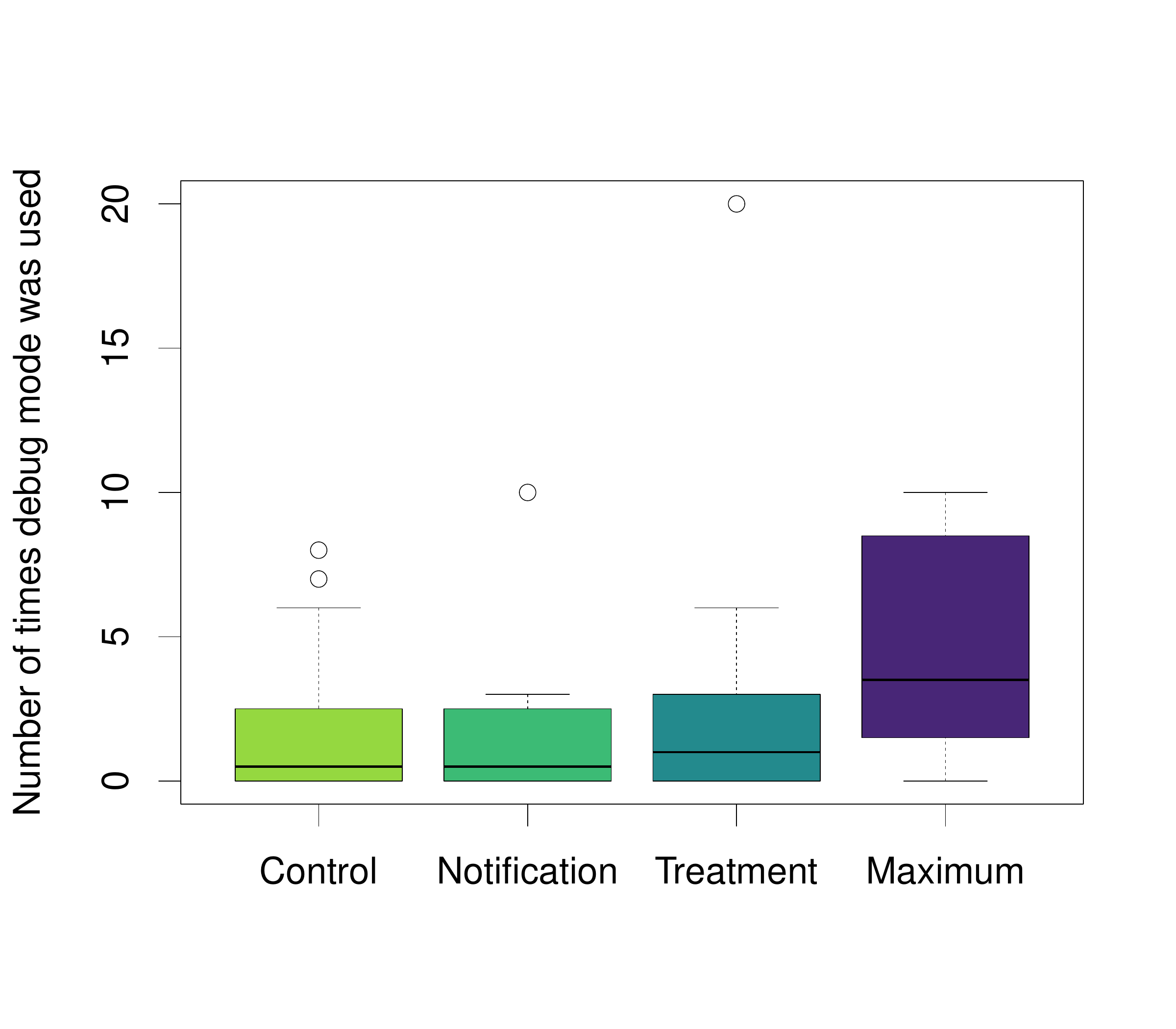}
			\vspace{-2.5em}
			\captionsetup{justification=centering}
			\caption{Number of times the IntelliJ \\debug mode was used}
			\label{fig:debugmode}
		\end{subfigure}
		\hfill
		\begin{subfigure}[t]{0.245\textwidth}
			\centering
			\includegraphics[width=\textwidth]{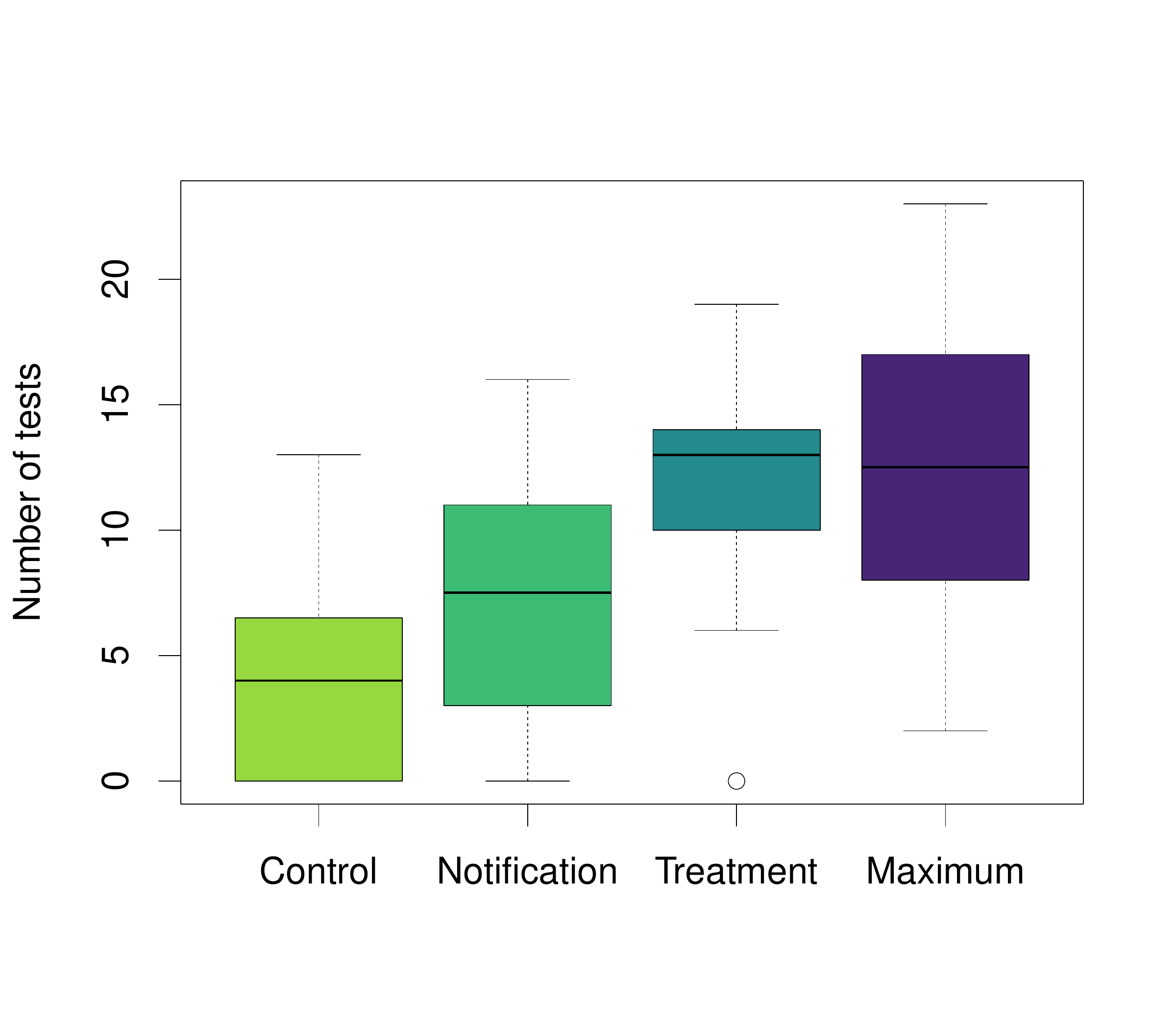}
			\vspace{-2.5em}
			\captionsetup{justification=centering}
			\caption{Number of tests written \\by the participants}
			\label{fig:testnumber}
		\end{subfigure}
		\hfill
		\begin{subfigure}[t]{0.245\textwidth}
			\centering
			\includegraphics[width=\textwidth]{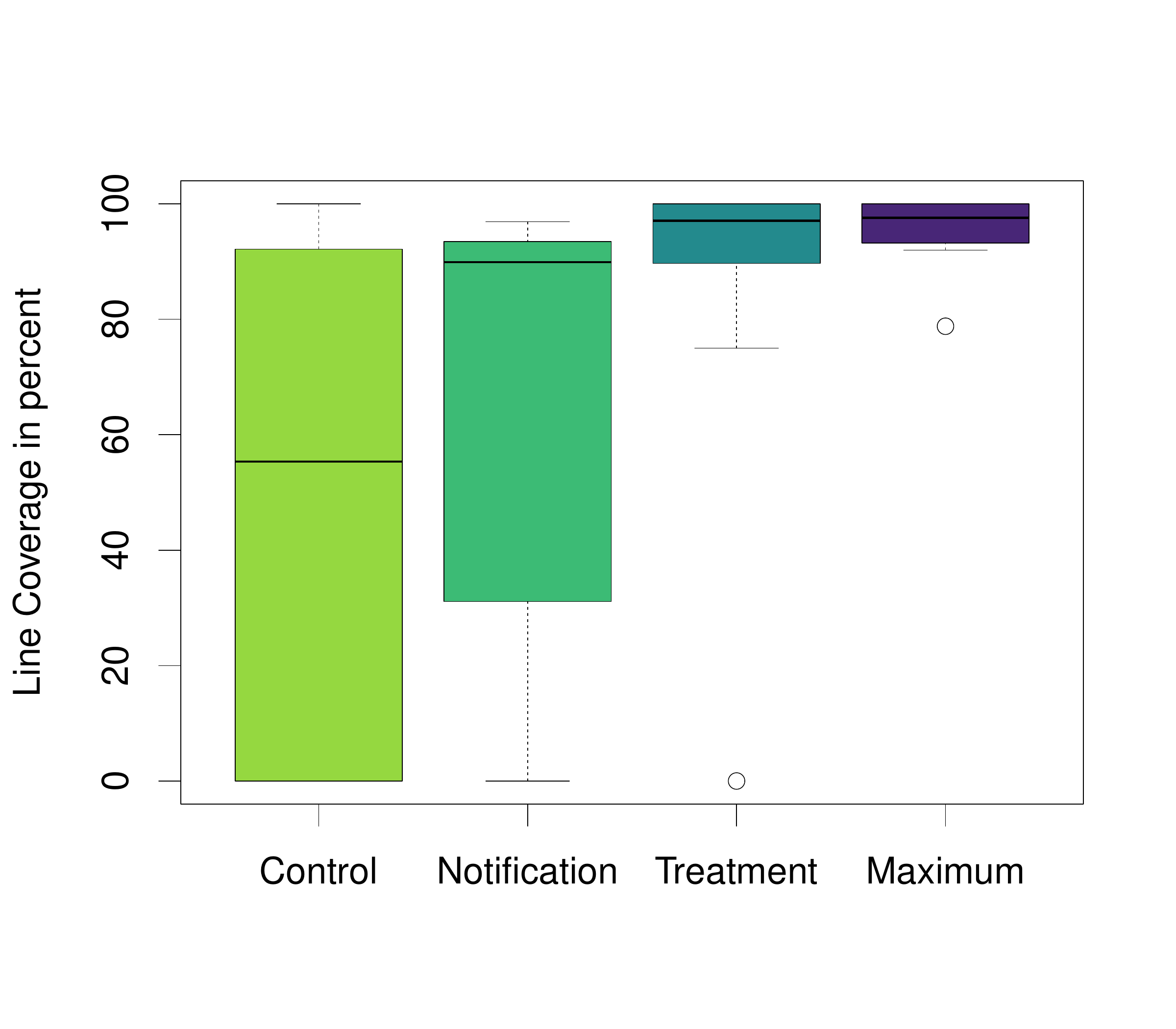}
			\vspace{-2.5em}
			\captionsetup{justification=centering}
			\caption{Line coverage measured \\by JaCoCo of the final test suites}
			\label{fig:linecoverage}
		\end{subfigure}
		\hfill
		\begin{subfigure}[t]{0.245\textwidth}
			\centering
			\includegraphics[width=\textwidth]{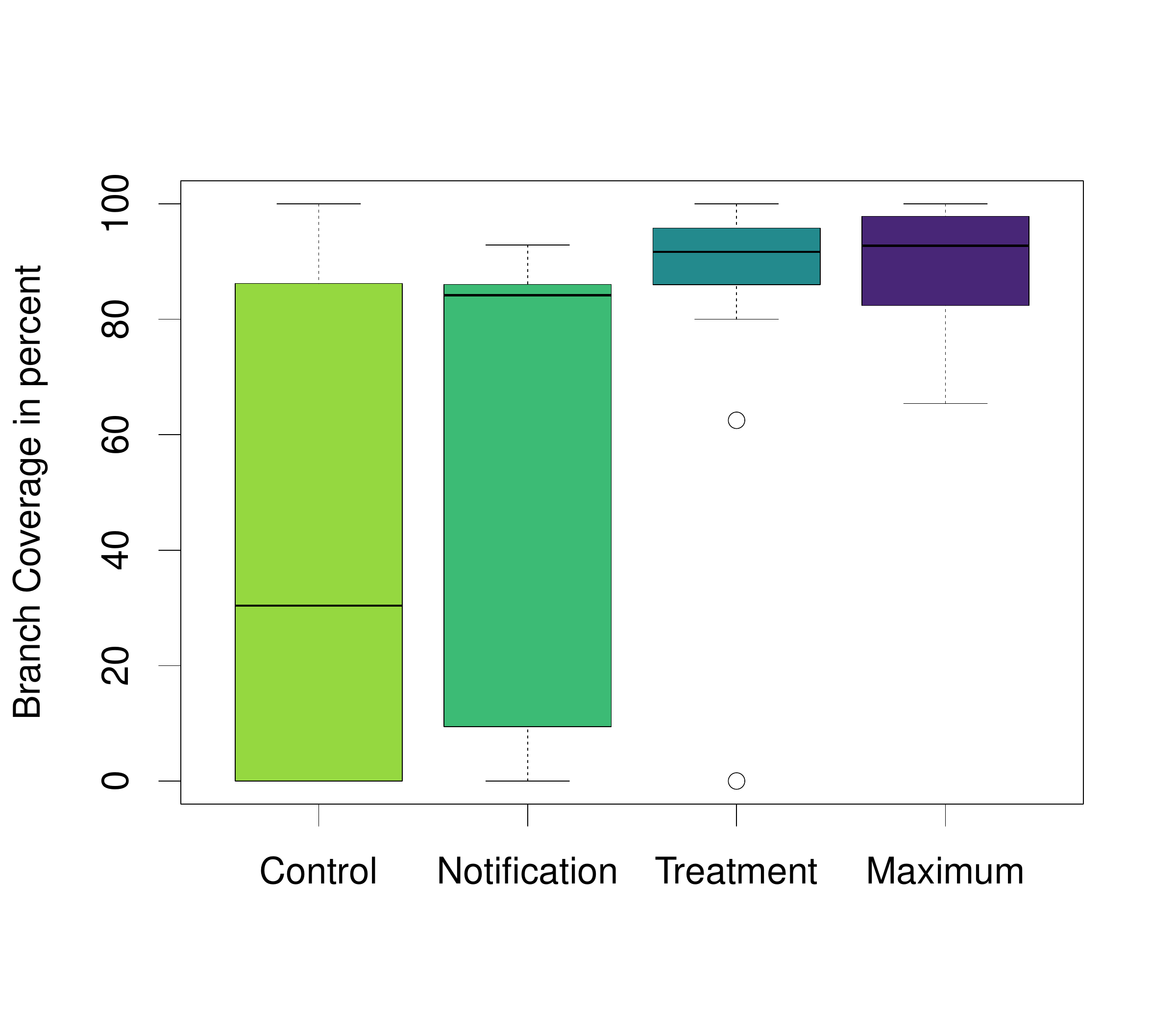}
			\vspace{-2.5em}
			\captionsetup{justification=centering}
			\caption{Branch coverage measured \\by JaCoCo of the final test suites}
			\label{fig:branchcoverage}
		\end{subfigure}
		\hfill
		\begin{subfigure}[t]{0.245\textwidth}
			\centering
			\includegraphics[width=\textwidth]{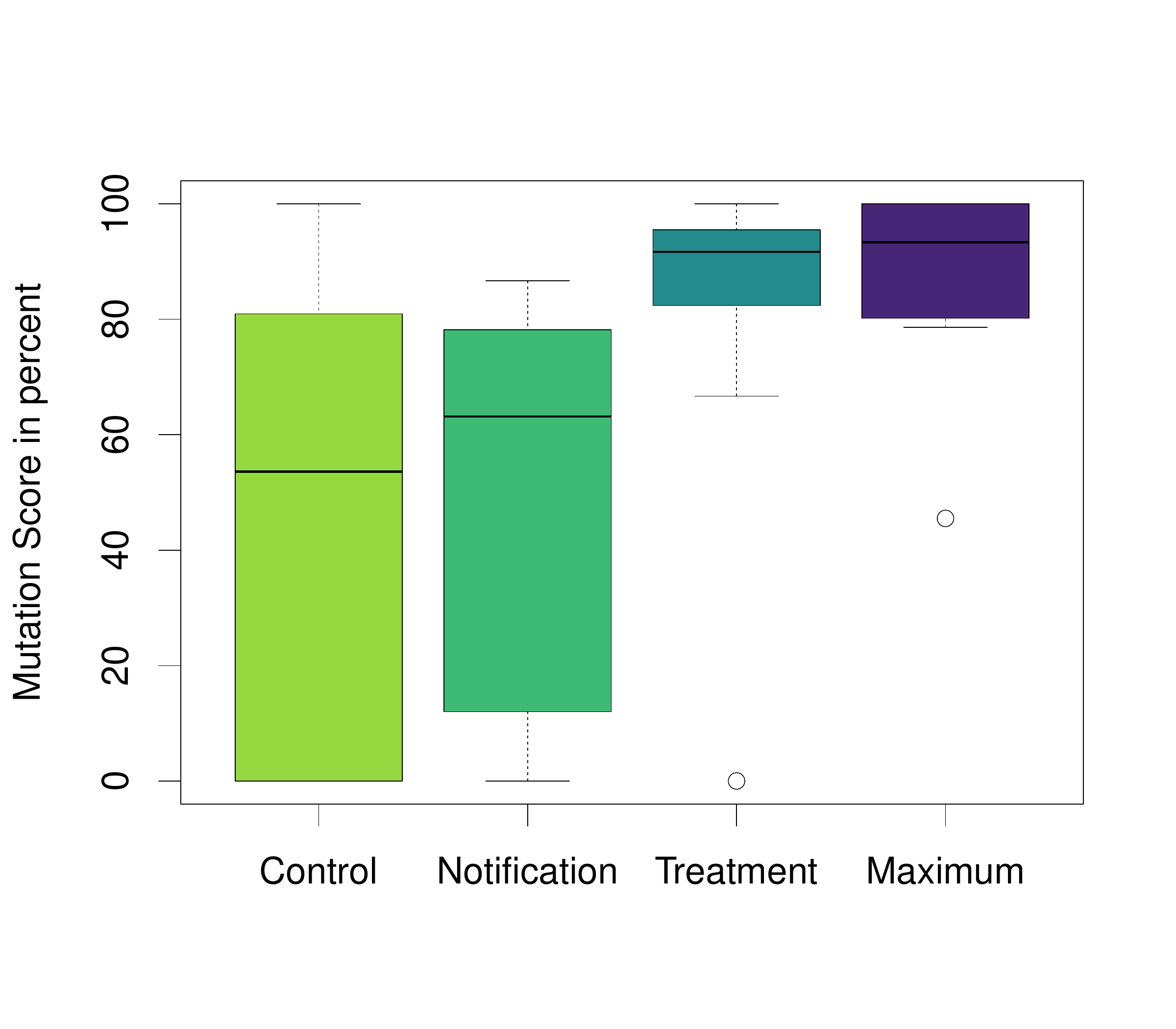}
			\vspace{-2.5em}
			\captionsetup{justification=centering}
			\caption{Mutation score measured \\by PITest of the final test suites}
			\label{fig:mutationscore}
		\end{subfigure}
		\hfill
		\begin{subfigure}[t]{0.245\textwidth}
			\centering
			\includegraphics[width=\textwidth]{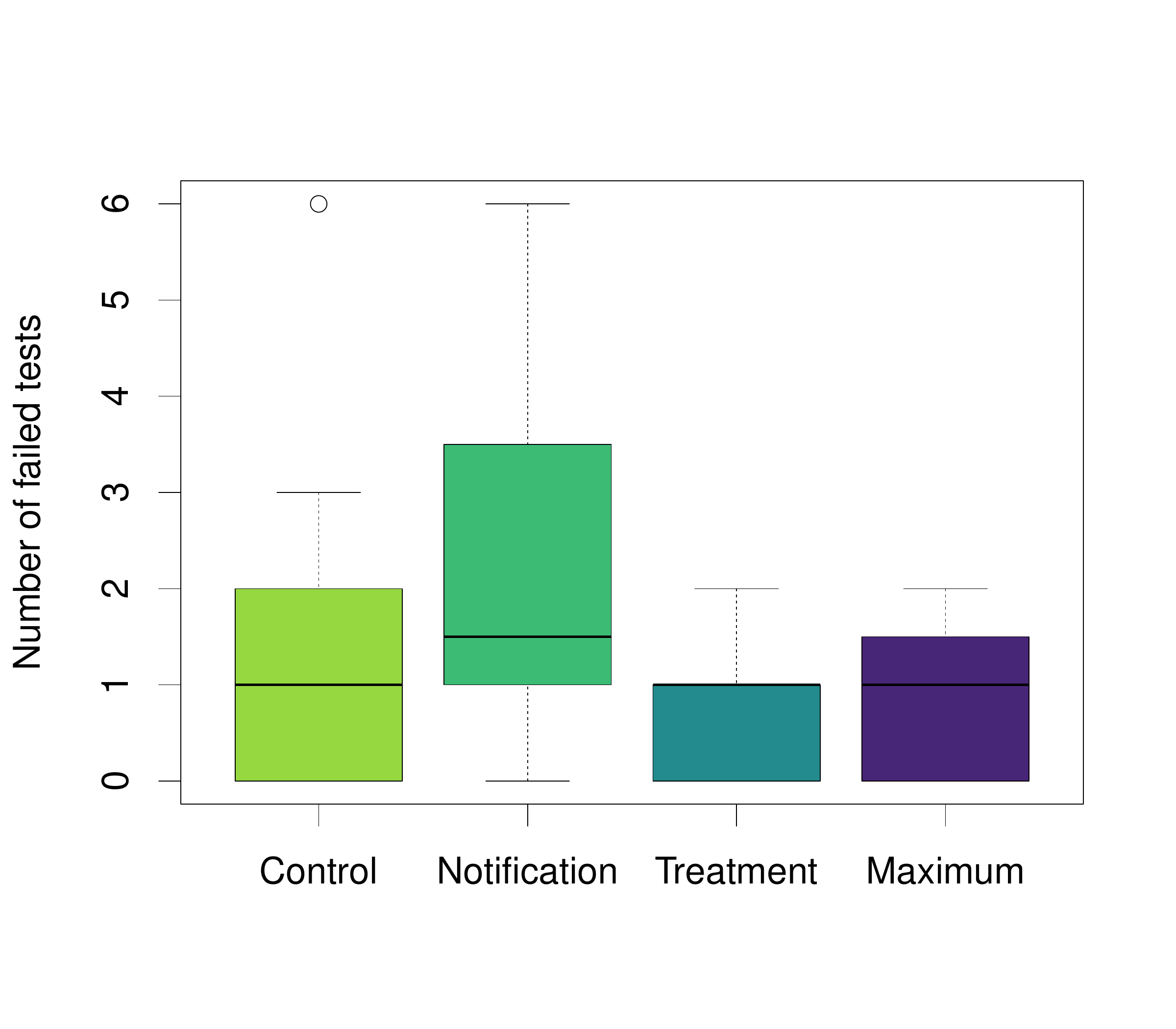}
			\vspace{-2.5em}
			\captionsetup{justification=centering}
			\caption{Number of failed tests accord- ing to the golden test suite}
			\label{fig:testsfailed}
		\end{subfigure}
		
		\caption{Differences between \control, \notifications, \treatment and \maximizing groups}
		\label{fig:boxplots}
	\end{figure*}
	
	\Cref{fig:testexecutions} shows the number of test executions,
        with a mean of 22.8 for the \treatment group and 10.5 for the
        control group. The \notifications ($p<0.01$), \treatment
        ($p<0.01$) and \maximizing ($p<0.01$) groups all executed
        significantly more tests than the \control group.  In the
        \maximizing group, two values with more than 1.5 million and
        250,000 test executions (achieved using IntelliJ's feature of
        repeating test executions a configurable number of times) were
        omitted from the graph to make it more readable. These two
        data points demonstrate the possible exploitation of \toolname
        when the participants try to get as many achievements as
        possible, which may distract the developers from their actual
        goal. Consequently, it seems important not to make the use of
        \toolname mandatory for developers, but rather to let them
        decide on their own how to use achievements as was done in the
        \treatment group.
        %
	%Therefore,
	%countermeasure have to be taken like a maximum progress per hour for specific achievements.
	%
	In addition, \cref{fig:testexecutionstime} illustrates the test
	executions over time together with the 84.6~\% confidence intervals. A
	significant difference between the \control and \treatment groups can be observed starting from minute 27.
	The \notifications and \control groups behave similarly to each other,
	while the \maximizing group shows a significant difference to the \control group in the end.
	%(which would be even more pronounced if including the two outlier participants mentioned above).
	Overall, participants with \toolname run their tests more often. % than the other groups.
	
	%In addition to the total number of test executions, the number of
	%executions over time can be analyzed with the help of the exact
	%Wilcoxon-Mann-Whitney test \cite{10.1214/aoms/1177730491} to determine
	%a significant difference between both groups. The results of this test
	%are visualized in 

	\paragraph{Coverage Measurement}
	
	\Cref{fig:testexecutionscoverage} compares the number of times
	tests were executed with coverage collection activated: Only
	one participant in the \control group used this functionality
	of IntelliJ 10~times, resulting in an average number of
	coverage executions of 0.6 overall participants. On the other
	hand, the \treatment group used the coverage report with a
	mean of 4.7 executions per participant.
	\Cref{fig:testexecutionscoveragetime} shows the coverage
	executions over time, demonstrating that the first participant
	of the \treatment group used the coverage report after 13
	minutes, whereas in the \control group only after 35
	minutes. This also means that the difference is significant
	from the beginning on ($p<0.01$), when the first
	participant executed their tests with coverage
		collection activated. The \notifications group behaves
	similarly to the \control group, even though there is a
	dedicated notification that suggests running tests with
	coverage. However, this is only shown if tests are run without
	coverage.
	The \maximizing group as expected measures coverage the most
	during the last third except the very end. Overall, the
	participants of the \treatment and \maximizing groups
	seem to care more about coverage.
	%since they
	%execute their tests more often with coverage than the \control and \notifications groups.

	\paragraph{Debug Executions}
	
	\Cref{fig:debugmode} shows how often tests were executed using
	the debug mode: On average participants of the \treatment
	group used debug executions 2.6 times, and of the \control
	group 1.8 times, although the difference is not significant
	($p=0.82$). This can also be seen in \cref{fig:debugtime},
	since the confidence intervals of all groups except for the
	\maximizing group overlap. We can see a significant difference
	halfway through the experiment, but the overall difference is
	not significant ($p=0.074$). We assume that debugging will be
	more common during maintenance and bug fixing, whereas our
	experiment task was to write new code. Furthermore, since the class under test is quite simple, the
		participants may not have had to use the debug mode to locate bugs
		found by failed tests.
	%One explanation could be that the participants with the enabled
	%plugin can delimit their errors because their tests more closely to
	%set breakpoints and use the debug mode the find faulty code.

	\summary{RQ 1}{\toolname influences the testing
		behavior of the participants: They use JUnit tests rather than
		main testing, create more tests, run their tests
		significantly more often, and use coverage reports more
		frequently. }
	%In addition, they are using the debug mode more often
	%  and are testing their code less often manually but with tests.}
	
	\subsection{RQ 2: Does \toolname influence resulting test suites?}
	
	\paragraph{Number of tests}
	
	\Cref{fig:testnumber} compares the resulting test suites in terms of
	the number of JUnit tests they consist of. The participants of the
	\treatment group wrote a mean of 11.5 tests, while the
	\control group wrote 4.3 tests and the \maximizing group 12.5. According to the
	exact Wilcoxon-Mann-Whitney test, the \treatment ($p<0.01$) and \maximizing ($p=0.036$) groups
	wrote significantly more tests than the \control group. %, while the number of tests per participant has a higher variation in the latter.
	One can also see that the \notifications group has more tests in their test suites (mean 7.38) than
		the \control group (mean 4.31), which again confirms that notifications influence developers to write more tests.
	
	\paragraph{Code coverage}
	
	\Cref{fig:linecoverage} compares the line coverage between the four
	groups, showing that the \treatment ($p<0.01$) and \maximizing ($p<0.01$) groups have significantly higher line
	coverage than the \control group. On average, the \treatment and \control groups achieve 89\,\% and 48\,\% line coverage, respectively. In the \treatment group, 
	there is one outlier without any coverage; this participant did not write any unit tests but only tested manually. When asked about this peculiar behavior, the participant stated to have rushed to complete and leave the study as quickly as possible.
	%The reason may be that the participant wanted to finish the task as fast as possible, because 
	%the participant also left the study early.
	%			
	Results are similar for branch coverage (\cref{fig:branchcoverage}),
	which also shows a significant difference between the control, \treatment ($p<0.01$) and \maximizing ($p<0.01$) groups, with a mean branch coverage of 85\,\% and 41\,\%
	for the \treatment and \control groups.
	The \notifications group has no significantly different line ($p=0.44$) or branch coverage ($p=0.64$) compared to the
	\control group, which is plausible since notifications only suggest measuring coverage, but not maximizing it.
	Overall, the significantly higher coverage for both the \treatment and \maximizing groups confirms the influence of gamification on test suite quality as measured with code coverage.
	
	\paragraph{Mutation score}
	
	The mean mutation score (\cref{fig:mutationscore}) of the \treatment
	group is 84\,\%, while the \control group achieves a
	mean score of 44\,\%, which is significantly lower
	($p<0.01$).
	The \notifications group shows no significant differences ($p=0.76$) compared to the \control group, again confirming the influence of gamification; the \maximizing group even achieves slightly higher mutation scores ($p=0.01$) than the \treatment group.
	%Therefore,
	%using the plugin results in higher mutation scores while notifications as a
	%stand-alone feature does not seem to lead to higher mutation scores.
	
	\summary{RQ 2}{\toolname has a significant influence on
		the resulting test suites, increasing the number of tests as well as the
		code coverage and mutation scores.}
	
	\subsection{RQ 3: Do achievement levels reflect differences in test suites and activities?}
	
	\begin{figure}
		\centering
		\begin{subfigure}[b]{0.48\linewidth}
			\centering
			\includegraphics[width=\linewidth]{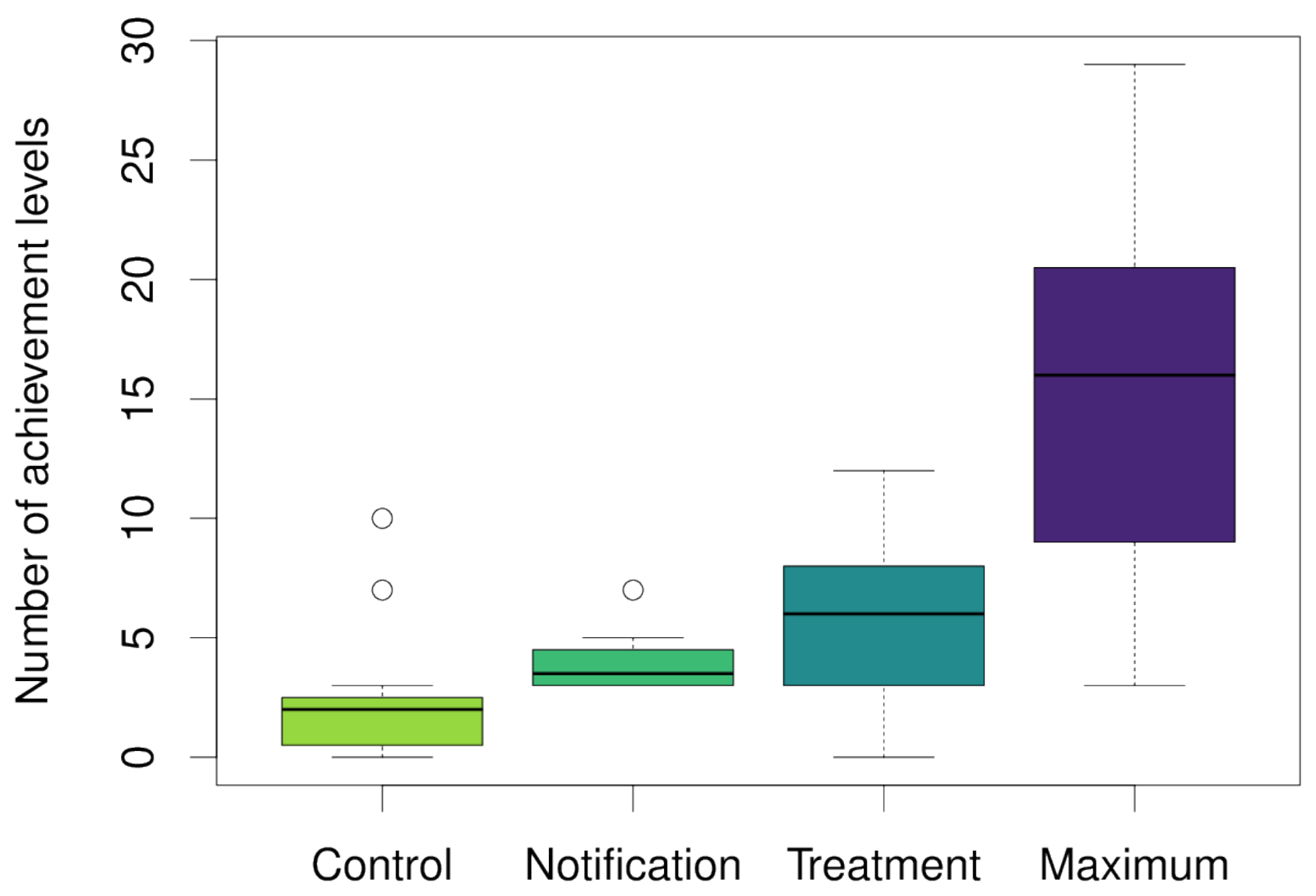}
			\captionsetup{justification=centering}
			\caption{The number of achievement levels reached}
			\label{fig:levels}
		\end{subfigure}
		\hfill
		\begin{subtable}[b]{0.48\linewidth}
			\small
			\setlength{\tabcolsep}{3pt}
			\begin{tabularx}{\linewidth}{lll}
				\toprule
				\textbf{Variable} & $\bm{r^2}$ & \textbf{p-value}   \\ \midrule
				Line Coverage & 0.21 & < 0.001 \\
				Branch Coverage & 0.23 & < 0.001 \\
				Mutation Score & 0.22 & < 0.001 \\
				Number of Tests & 0.39 & < 0.001 \\
				\bottomrule
			\end{tabularx}%
			\setlength{\tabcolsep}{6pt}
			\captionsetup{justification=centering}
			\caption{The Pearson correlations based on the number of levels}
			\label{tab:corr}
		\end{subtable}
		
		\caption{The number of achievement levels and their Pearson correlations with different test suite metrics}
		\label{fig:levcorr}
	\end{figure}
	
	RQ1 demonstrated that participants of the \maximizing group
	engage most with testing activities, followed by the
	\treatment group, then the \notifications group, and finally
	the \control group, who motivated the least with testing.
	\Cref{fig:levels} compares how many achievement levels the
	participants of the four groups reached, and confirms this
	ranking of the groups with a mean of 2.6 for the \control group, 4
	for the \notifications group, 6.1 for the \treatment group, and 15.4 for
	the \maximizing group; \cref{fig:levelstime} shows that this holds throughout the entire duration of the experiment.   All groups reached significantly more
	levels than the \control group (\notifications $p<0.01$,
	\treatment $p<0.01$, \maximizing $p<0.01$).
	Thus, developers who have been awarded achievements indeed are more committed to testing.
	To see whether developers with more achievements and higher achievement levels also
		produce overall better test suites, \cref{tab:corr} shows the
		Pearson rank correlations between test suite metrics (RQ2)
		and achievement levels. There is a significant moderate correlation between the number of tests and achievements and a significant weak positive correlation for all other metrics. This is a strong affirmation of the gamification approach, as it shows that acquiring achievements is indeed related to producing better test suites.
	
	%This shows that both the 
	%notifications as stand-alone feature and the whole plugin with achievements 
	%motivates the participants to test enough to reach multiple levels. 
	%
	%The \control group has two outliers,
	%who reached seven and ten levels, which leads to the conclusion that they
	%cared about testing their code even without \toolname in contrast to the 
	%other participants of the \control group. 
	%
	%When comparing the number of tests written by the participants (\cref{fig:testnumber})
	%with the achievement levels (\cref{fig:levels}), one can see clear similarities between 
	%these, which leads to the conclusion that the more they tested the more levels they reached.
	
	\summary{RQ 3}{Higher achievement levels are indicators of better motivation with testing and better resulting test suites.}

	\subsection{RQ 4: Does \toolname influence the functionality of resulting code?}
	
	\Cref{fig:passedteststime} shows the number of passed tests of the golden test suite during the experiment, where notably the \treatment and \maximizing groups exhibit passing tests earlier compared to the other groups. Particularly, the \maximizing group reached a plateau where most tests passed at minute 43, indicating that solving achievements leads to faster identification and resolution of program faults as well as faster implementation of new functionality.
	
	At the end of the experiment, the average number of failing tests of the golden test suite (\cref{fig:testsfailed}) is 0.9 for the \treatment group, and 1.6 for the \control group, with no statistically significant ($p=0.45$) difference. The \maximizing and \notifications groups also show similar results to the other groups, although participants in the notifications group have the highest average number of failing tests (2.25). The lack of significant differences can be attributed to several factors: (1) the experiment duration was long enough for participants to complete the implementation even with inferior testing behavior, (2) the golden test suite consisted of only six test cases, and (3) the two methods to be implemented allowed for a limited number of distinct possible bugs.
	
	\summary{RQ 4}{The achievements of \toolname lead to earlier implementation of functionality, although we observed no significant differences in functionality at the end of the experiment.}
	
	\subsection{RQ 5: Does \toolname influence the developer experience?} \label{sec:rq4exit}

	\begin{figure}
		\includegraphics[width=\linewidth]{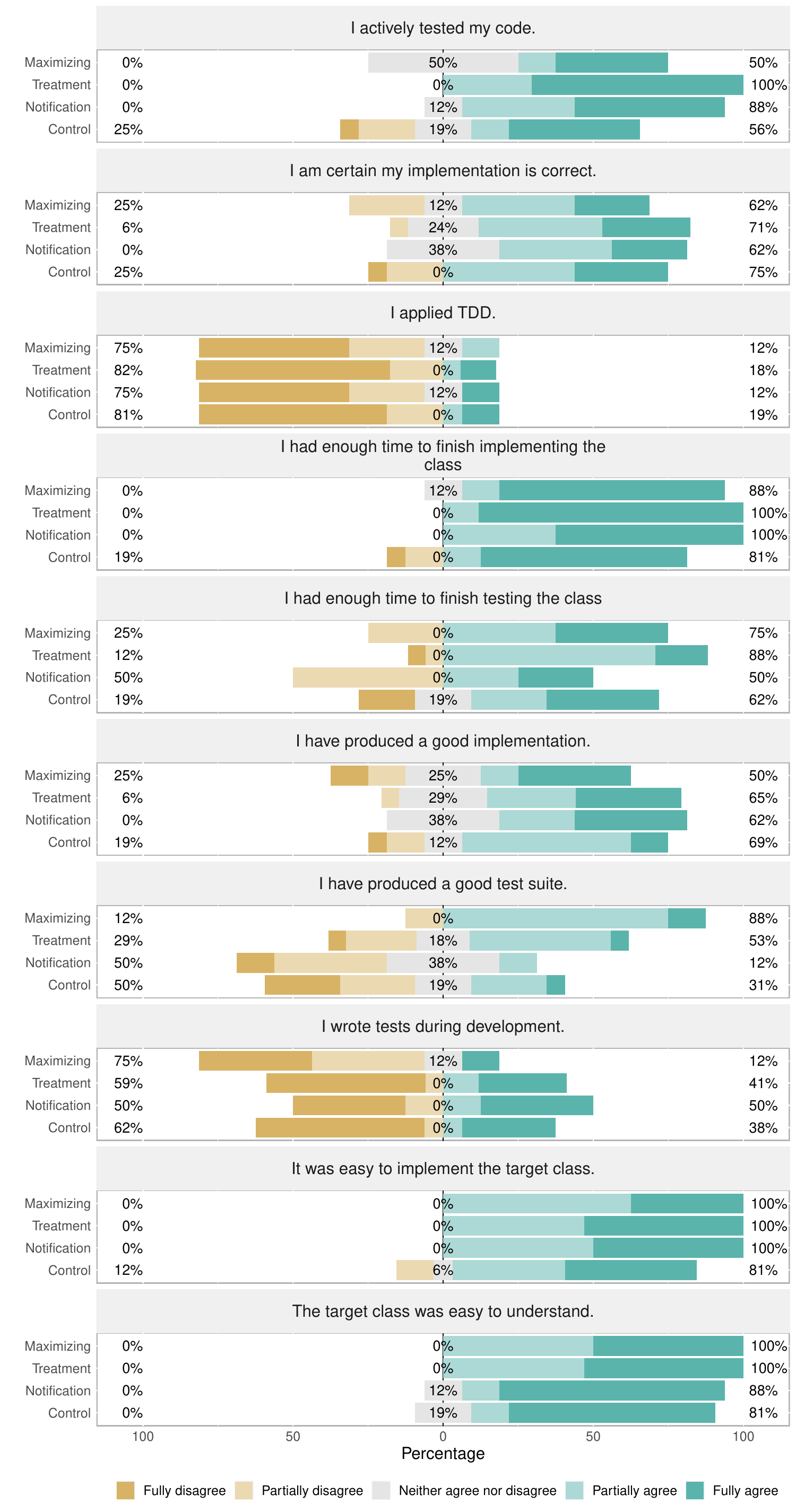}
		\vspace{-1em}
		\caption{General survey responses}
		\label{fig:exitall}
	\end{figure}

	\begin{figure}
		\includegraphics[width=\linewidth]{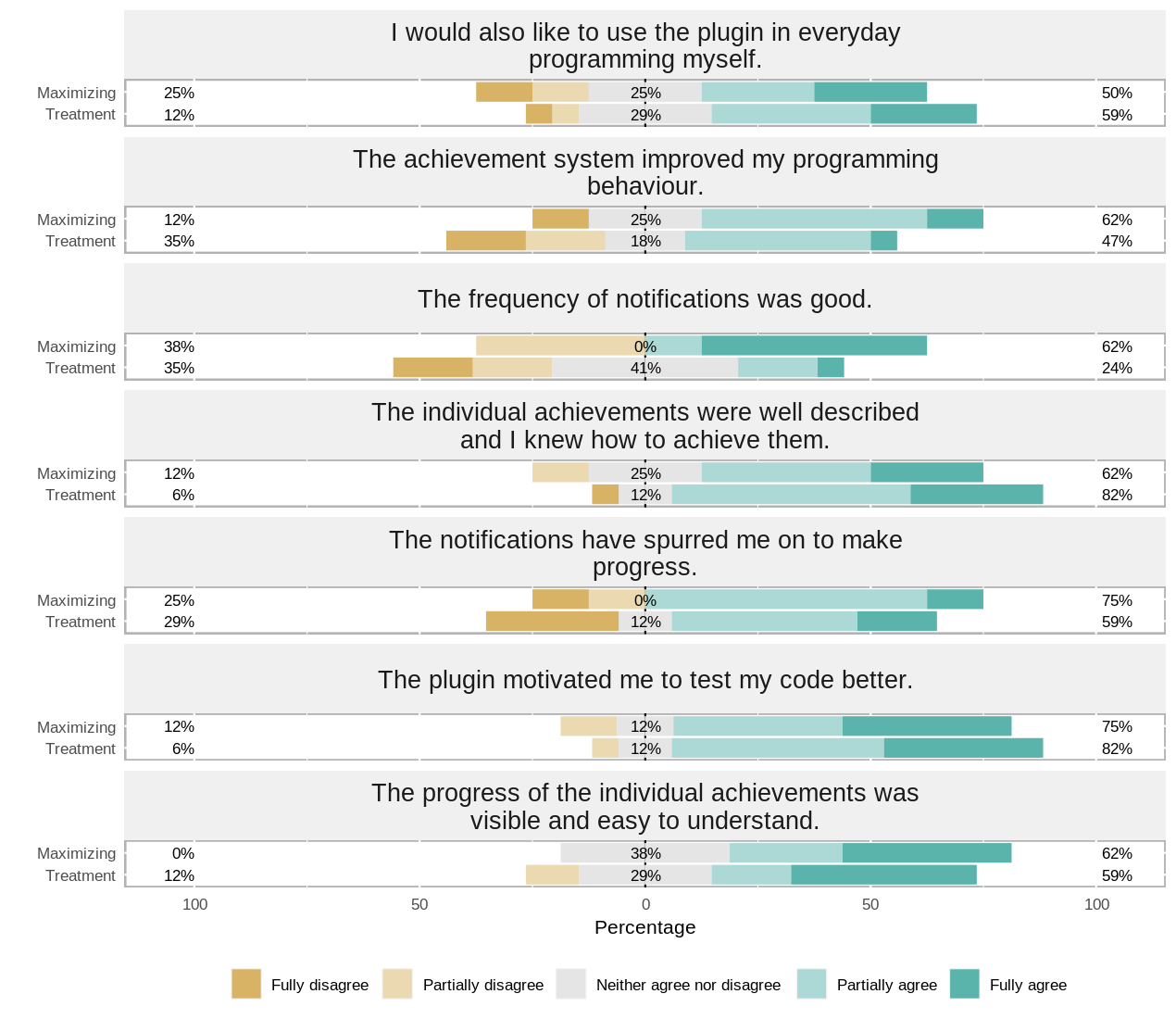}
		\vspace{-1em}
		\caption{Answers regarding the \toolname plugin}
		\label{fig:exitplugin}
	\end{figure}

	%At the end of the experiment, all participants were asked to fill out
	%an exit survey about their work done (\cref{sec:exitsurvey}).
	
	According to the exit survey (\cref{fig:exitall}), the target class was well chosen. All groups claim that they had enough time to implement and test the class and that it was easy to understand and implement.
	%There are some differences between the \treatment and \control groups visible in the data since more participants of the \control group disagreed on whether the class was easy to work with. \todo{Check for significance here also} The reasons may be that the \control group found bugs during main testing and spent a lot of time fixing them without being able to automatically verify the result with a good test suite.
	%
	Notable and significant differences can be observed on whether
	participants believe to have tested their code actively: All
	participants of the \treatment group claim to have actively
	tested their code, while only 55\,\% of the \control group did
	this; this difference is significant ($p=0.03$). Participants
	of the \treatment group are also more confident in that they
	produced a good test suite than participants of the \control
	group, although the difference is not significant
	($p=0.16$). Participants of the \maximizing group, who
	invested more time into testing, believe even more often to
	have produced a good test suite, significantly so when
	compared to the \notifications ($p=0.01$) and \control
	($p=0.025$) groups, but not against the \treatment group
	($p=0.14$). Throughout the survey, we generally observe
	favorable responses by the \treatment group, although not
	statistically significant for the other
	cases. Interestingly, members of the \maximizing group are
	least confident in that they produced a good implementation,
	which indicates a disadvantage of their focus on the
	achievements rather than the \mbox{programming task.}

	%	\subsection{RQ 5: Did the plugin motivate the participants to write more tests?}
	Considering the experiences of the \treatment and \maximizing
	groups, who both used achievements (\cref{fig:exitplugin}),
	more than half the participants perceived that \toolname
	improved their programming behavior, and more than 60\,\% were
	driven by notifications, including both encouraging and
	progress notifications. Most of the participants understood
	the individual achievements and how to achieve them. Nearly
	60\,\% would like to use \toolname in their everyday work.
	%,
	%while less than 20\,\% would not use it.
	More than 75\,\% of the participants claim to have been
	motivated to improve testing, which
	indicates that the plugin served its purpose.
	
	\summary{RQ 5}{Users of \toolname perceived the task as easier, are more confident about their results, and confirm that the achievements motivated them to apply more testing.}

%% file: sections/conclusion.tex
	\toolname adds test achievements to the IntelliJ IDE. In an experiment with 49 participants, we found that these achievements significantly improved developers' behavior with respect to writing tests, running and using these tests, and analyzing them. We believe this approach is especially applicable in short-term scenarios like onboarding new staff, fostering a better testing culture in existing teams, or introducing new techniques, since \toolname only needs to be used until testing practices are understood or adopted.
	Nevertheless it is similarly conceivable that test achievements become a permanent part of developers' coding environment. However, further research is required to assess the long-term effectiveness of \toolname and the potential to inspire engagement in developers, as player retention likely becomes a challenge~\cite{DBLP:conf/icsoc/FatimaAMU21}.

	There are ample opportunities for further improving \toolname or adding new achievements. For example, our study participants provided suggestions such as rewarding early testing to promote Test-Driven Development, or adapting achievements to individual developers by adjusting achievement levels based on the currently opened project in the IDE or a developer's performance at coding or testing. Furthermore, there is potential to include new game elements and foster competition among developers, such as through a leaderboard---which may be particularly useful when aiming at long-term use of gamification.

	%
	%Future work could also replicate our study with different programming
	%languages to validate and apply our findings in real-world
	%scenarios.
	In order to support experiment replications and further
        research on gamifying developer behavior in the IDE, all our
        experiment material is available at:
        \begin{center}
          \url{https://doi.org/10.6084/m9.figshare.22353352.v1}
        \end{center}
	The source code of \toolname is available at:
        \begin{center}
          \url{https://github.com/se2p/IntelliGame}
        \end{center}